\documentclass[10pt,journal,cspaper,compsoc]{IEEEtran}
\usepackage{amsmath}
\usepackage{graphicx,epsfig,color,endnotes,alltt}
\usepackage{caption}
\usepackage{subcaption}
\usepackage{listings}
\usepackage{url}
\usepackage{balance}
\usepackage{algorithm}
\usepackage[noend]{algorithmic}
\usepackage{fancyvrb}
\usepackage{bbding}
\usepackage[normalem]{ulem}
\usepackage{array}
\usepackage{color}
\usepackage{multirow}
\usepackage{cite}
\usepackage{fmtcount}
\usepackage{epstopdf}
\usepackage{slashbox}

\newenvironment{frcseries}{\fontfamily{frc}\selectfont}{}
\newcommand{\textfrc}[1]{{\frcseries#1}}

\definecolor{Magenta}{rgb}{1,0.5,0}

\newcommand{\jianlong}[1]{{#1}}

\ifCLASSOPTIONcompsoc

\else
\fi

\ifCLASSINFOpdf
\else
\fi

\begin{document}
%
\title{Kernelet: High-Throughput GPU Kernel Executions with Dynamic Slicing and Scheduling}
\author{Jianlong~Zhong,
        Bingsheng~He
\IEEEcompsocitemizethanks{\IEEEcompsocthanksitem J. Zhong and B. He
are with School of Computer Engineering, Nanyang Technological
University, Singapore,
639798.\protect\\
E-mail: jzhong2@ntu.edu.sg, bshe@ntu.edu.sg}
\thanks{}}

\IEEEcompsoctitleabstractindextext{%
\begin{abstract}

Graphics processors, or GPUs, have recently been widely used as
accelerators in the shared environments such as clusters and clouds.
In such shared environments, many kernels are submitted to GPUs from
different users, and throughput is an important metric for
performance and total ownership cost. Despite the recently improved
runtime support for concurrent GPU kernel executions, the GPU can be
severely underutilized, resulting in suboptimal throughput. In this
paper, we propose \emph{Kernelet}, a runtime system with dynamic
slicing and scheduling techniques to improve the throughput of
concurrent kernel executions on the GPU. With slicing, Kernelet
divides a GPU kernel into multiple sub-kernels (namely
\emph{slices}). \jianlong{Each slice has tunable occupancy to allow
co-scheduling with other slices and to fully utilize the GPU
resources.} We develop a novel and effective Markov chain based
performance model to guide the scheduling decision. Our experimental
results demonstrate up to 31.1\% and 23.4\% performance improvement
on NVIDIA Tesla C2050 and GTX680 GPUs, respectively.
\end{abstract}

\begin{keywords}
Graphics processors, Dynamic scheduling, Concurrent kernel
executions, Dynamic Slicing, Performance models
\end{keywords}}

\maketitle

\IEEEdisplaynotcompsoctitleabstractindextext

\IEEEpeerreviewmaketitle

\section{Introduction}\label{sec:intro}
\jianlong{The graphics processing unit (or GPU) has} become an
effective accelerator for a wide range of applications from
computation-intensive applications (e.g.,
~\cite{Volkov:2008:BGT:1413370.1413402,Nath:2011:OSD:2063384.2063392,SC08_2})
to data-intensive applications (e.g, ~\cite{NagaTerasort, gpujoin}).
Compared with multicore CPUs, new-generation GPUs can have much
higher computation power in terms of \jianlong{FLOPS} and memory
bandwidth. For example, an NVIDIA Tesla C2050 GPU can deliver the
peak single precision floating point performance of over one Tera
FLOPS, and memory bandwidth of 144 GB/s. Due to their
\jianlong{immense} computation power and memory bandwidth, GPUs have
been integrated into cluster\jianlong{s} and cloud computing
infrastructures. In Top500 list of November 2012, two out of the top
ten {supercomputers} are with GPUs integrated. Amazon and Penguin
have provided virtual machines with GPUs. In both cluster and cloud
environments, GPUs are often shared by many concurrent GPU programs
(or \emph{kernels}) (most likely submitted by multiple users).
Additionally, to enable sharing GPUs remotely, a number of software
frameworks such as rCUDA~\cite{rcuda} and V-GPU~\cite{vgpu} have
been developed. This paper studies whether and how we can improve
the throughput of concurrent kernel executions on the GPU.

Throughput is an important optimization metric \jianlong{for
efficiency} and the total ownership cost of GPUs in such shared
environments. First, many GPGPU applications such as scientific and
financial computing tasks are usually throughput
oriented~\cite{garlandThroughputOriented}. A high throughput leads
to the high performance and productivity for users. Second, compared
with CPUs, GPUs are still expensive devices. Therefore, a high
throughput not only means a high utilization on GPU resources but
also the total ownership cost of running the application on the GPU.
That might also be one of the reasons that GPUs are usually deployed
and shared to handle many kernels from users.

Recently, we have witnessed the success of GPGPU research. However,
most studies focus on single-kernel optimizations (e.g., new data
structures~\cite{hetods} and GPU friendly computing
patterns~\cite{NagaTerasort, NagaFFT}). Despite the fruitful
research, a single kernel usually severely under-utilizes the GPU.
This severe underutilization is mainly due to the inherent memory
and computation behavior of a single kernel (e.g., irregular memory
accesses and execution pipeline stalls). In our experiments, we have
studied eight common kernels (details are presented in
Section~\ref{sec:evaluation}). On C2050, their average
$\mathit{IPC}$ is 0.52, which is far from the optimal value (1.0).
Their memory bandwidth utilization is only ranging from 0.02\% to
14\%.

Recent GPU architectures like NVIDIA Fermi~\cite{fermi_white_paper}
architecture supports concurrent kernel executions, which allows
multiple kernels to be executed on the GPU simultaneously if
resources are allowed. In particular, Fermi adopts a form of
\emph{cooperative} kernel scheduling. Other kernels requesting the
GPU must wait until the kernel occupying the GPU voluntarily yields
control. Here, we use NVIDIA CUDA's terminology, simply because CUDA
is nowadays widely adopted in GPGPU applications. A kernel consists
of multiple executions of \emph{thread blocks} with the same program
on different data, where the execution order of thread blocks is not
defined. \jianlong{On the Fermi GPUs}, one kernel can take the
entire GPU as long as it has sufficient thread blocks to occupy all
the multi-processors (even though it can have severely low resource
utilization). Concurrent execution of two such kernels almost
degrades to sequential execution on individual kernels. Recent
studies~\cite{gpu_consolidation_hpdc11} on scheduling the concurrent
kernels mainly focus on the kernels with low occupancy (i.e., the
thread blocks of a single kernel cannot fully utilize all GPU
multiprocessors). However, the occupancy of kernels (with large
input data sizes in practice) is usually high after single-kernel
optimizations.

Individual kernels as a whole cannot realize the real sharing of the
{GPU} resources. A natural question is: can we slice the kernel into
small pieces and then co-schedule slices from different kernels in
order to improve the {GPU} resource utilization? The answer is yes.
One observation is that {GPU} kernels (e.g., those written in {CUDA}
or {OpenCL}) conform to the {SPMD} (Single Program Multiple Data)
execution model. In such data-parallel executions, a kernel
execution can usually be divided into multiple \emph{slices}, each
consisting of multiple thread blocks. Slices can be viewed as
low-occupancy kernels and can be executed simultaneously with slices
from other kernels. The GPGPU concurrent kernel scheduling problem
is thus converted to the slice scheduling problem.

With slicing, we have two more issues \jianlong{to address}. The
first issue is on the slicing itself: what is the suitable slice
size? How to perform the slicing in a transparent manner? The
smallest granularity of \jianlong{a slice} is one thread block,
which can lead to significant runtime overhead of submitting too
many such small slices onto the {GPU}s for execution. To the other
extreme, the largest granularity of the slice is the entire kernel,
which degrades to the non-sliced execution. The second issue is how
to select the slices for co-schedule in order to maximize the {GPU}
utilization.

To address those issues, we develop \emph{Kernelet}, a runtime
system with dynamic slicing and scheduling techniques to improve the
GPU utilization. Targeting at the concurrent kernel executions in
the shared GPU environment, Kernelet dynamically performs slicing on
the kernels, and the slices are carefully designed with tunable
occupancy to allow slices from other kernels to utilize the GPU
resources in a complementary way. For example, one slice utilizes
the computation units and the other one on memory bandwidth. We
develop a novel and effective Markov chain based performance model
to guide kernel slicing and scheduling in order to maximize the GPU
resource utilization. Compared with existing GPU performance models
which are limited to a single kernel only, our model are designed to
handle heterogeneous workloads (i.e., slices from different
kernels). We further develop a greedy co-scheduling algorithm to
always co-schedule the slices from the two kernels with the highest
performance gain according to our performance model.

We have evaluated Kernelet on two latest GPU architectures (Tesla
C2050 and GTX680). The GPU kernels under study have different memory
and computational characteristics. Experimental results show that 1)
our analytical model can accurately capture the performance of
heterogeneous workloads on the GPU, 2) our scheduling increase the
GPU throughput by up to 31.1\% and 23.4\% on C2050 and GTX680,
respectively.

{\bf Organization.} The rest of the paper is organized as follows.
We introduce the background and definition of our problem in
Section~\ref{sec:motivation}. Section~\ref{sec:systemdesign}
presents the system overview, followed by detailed design and
implementation in Section~\ref{sec:methodology}. The experimental
results are presented in Section~\ref{sec:evaluation}. We review the
related work in Section~\ref{sec:related} and conclude this paper in
Section~\ref{sec:conclusion}.

\section{Background and Problem Definition}\label{sec:motivation}

In this section, we briefly introduce the background on GPU
architectures, and next present our problem definition.

\subsection{GPU Architectures}
GPUs have rapidly evolved into a powerful accelerator for many
applications, especially after CUDA was released by
NVIDIA~\cite{CUDA}. The top tags in http://gpgpu.org/ show that a
wide range of applications have been accelerated with GPGPU
techniques, including vision and graphics, image processing, linear
algebra, molecular dynamics, physics simulation and scientific
computing etc. Those applications cover quite a wide range of
computation and memory intensiveness. In the shared environment like
clusters and cloud, it is very likely that users submit kernels from
different applications to the same GPU. Thus, it is feasible to
schedule kernels with different memory and computation
characteristics to \jianlong{better} utilize GPU resources.

This paper focuses on the design and implementation with NVIDIA
CUDA. Since OpenCL and CUDA have very similar designs, our design
and implementation can be extended to OpenCL with little
modification. Kernelet takes advantage of concurrent kernel
execution capability of new-generation GPUs like NVIDIA Fermi GPUs.
With the introduction of CUDA, a GPU can be viewed as a many-core
processor with a set of streaming multi-processors (SM). Each SM has
a set of scalar cores, which executes the instructions in the SIMD
(Single Instruction Multiple Data) manner. The SMs are in turn
executed in the SPMD manner. The program is called \emph{kernel}.

In CUDA's abstraction, GPU threads are organized in a hierarchical
configuration: usually 32 threads are firstly grouped into
\jianlong{a warp}; warps are further grouped into thread blocks. The
CUDA runtime performs mapping and scheduling at the granularity of
thread blocks. Each thread block is mapped and scheduled on an SM,
and cannot be split among multiple SMs. Once a thread block is
scheduled, its warps become active on the SM. Warp is the smallest
scheduling unit on the GPU. \jianlong{Each SM uses one or more
\emph{warp schedulers} to issue instructions of the active warps.}
Another important resource of the GPU is shared memory, which is a
small piece of scratchpad memory at the scope of a thread block. It
is small and has very low latency. Shared memory is visible for all
the threads in the same thread block.

We define the SM occupancy as the ratio of active warps to the
maximum active warps that are allowed to run on the SM. Higher
occupancy means higher thread parallelism. The aggregated register
and shared memory usage of all warps should not exceed the total
amount of available registers and shared memory on an SM.

{\bf Block Scheduling.} CUDA runtime system maps thread blocks to
SMs in a round-robin manner. If the number of thread blocks in a
kernel is less than the number of SMs on the GPU, each thread block
will be executed on a dedicated SM; otherwise, multiple thread
blocks will be mapped to the same SM. The number of thread blocks
that can be \emph{concurrently} executed on the SM depends on their
total resource requirements (in terms of registers and shared
memory). If the kernel currently being executed on the GPU cannot
fully utilize the GPU, the GPU allows to schedule thread blocks from
other kernels for execution.

{\bf GPU Code Compilation.} In the process of CUDA compilation, the
compiler first compiles the CUDA C code to PTX (Parallel Thread
Execution) code. PTX is a virtual machine assembly language and
offers a stable programming model for evolving hardware
architectures. PTX code is further compiled to native GPU
instructions (SASS). The GPU can only execute the SASS code. CUDA
executables and libraries usually provide either PTX code or SASS
code or both. In the shared environments, the source code is usually
not available for kernel scheduling upon users submit the kernels.
Thus, Kernelet should be designed to work on both PTX and SASS code.

\subsection{Problem Definition}\label{sec:problemDefinition}

\begin{figure}
        \centering
        \begin{subfigure}{0.24\textwidth}
                \centering
                \includegraphics[width=\textwidth]{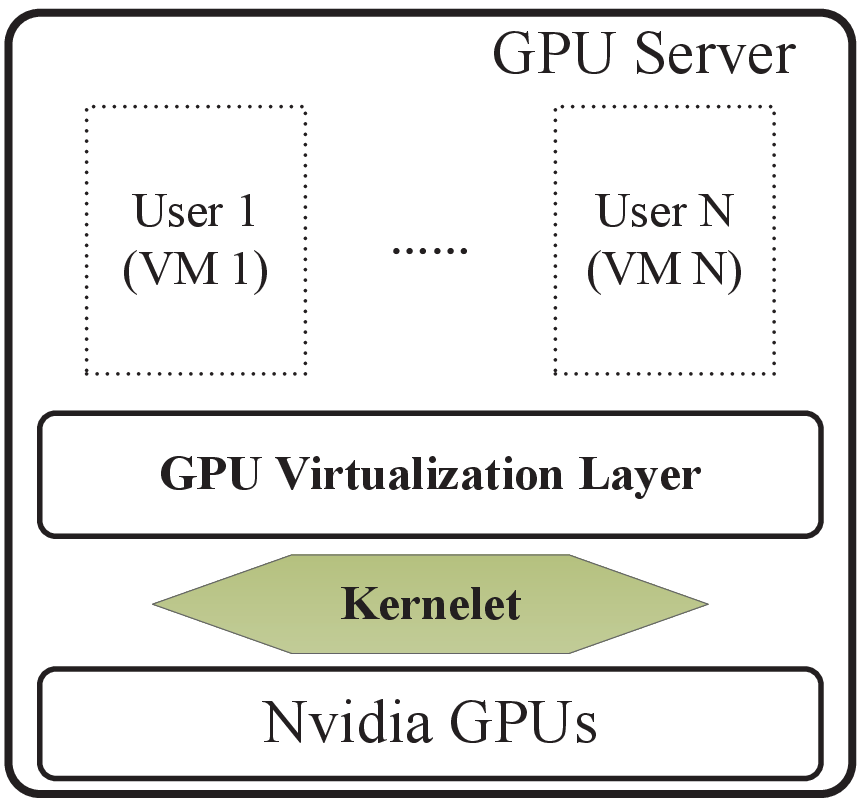}
                \caption{GPU shared within one machine.}
                \label{fig:scenario_vm}
        \end{subfigure}
        \begin{subfigure}{0.24\textwidth}
                \centering
                \includegraphics[width=\textwidth]{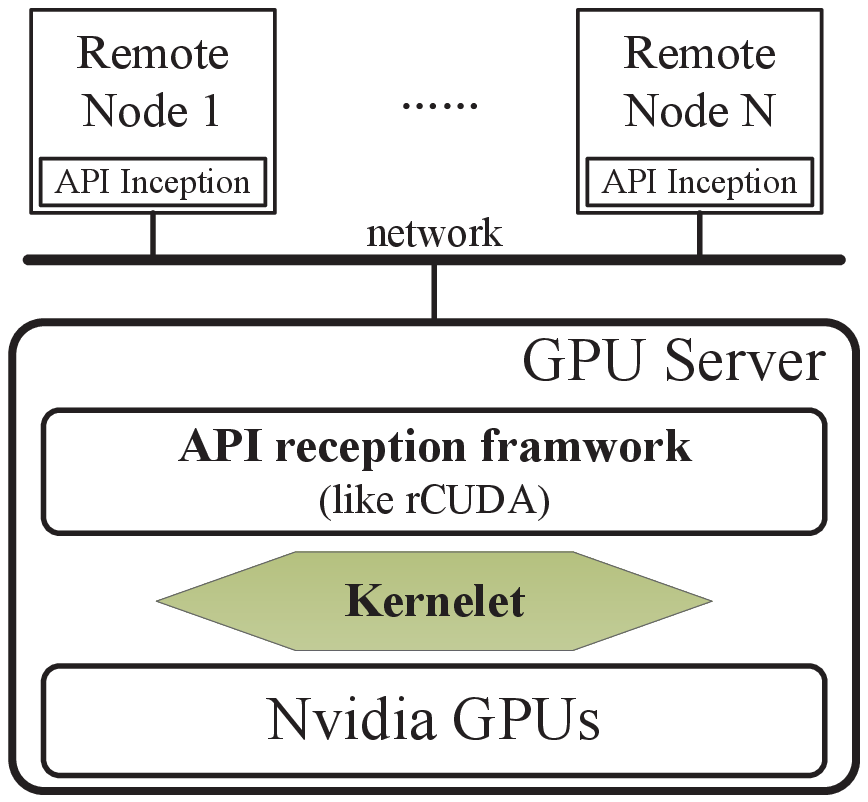}
                \caption{GPU shared by remote clients.}
                \label{fig:scenario_cluster}
        \end{subfigure}
        \caption{Application scenarios of concurrent kernel executions on the GPUs.}
        \label{fig:applicationScenario}
\end{figure}

{\bf Application scenario.} We consider two typical application
scenarios in the shared environments as shown in
Figure~\ref{fig:applicationScenario}. One is sharing the GPUs among
multiple tenants in the virtualized environment (e.g., cloud). As
illustrated in Figure~\ref{fig:scenario_vm}, there is usually a GPU
virtualization layer integrated with the hypervisor.
Figure~\ref{fig:scenario_cluster} shows the other scenario, in which
GPU servers offer API reception softwares (like rCUDA~\cite{rcuda,
vgpu}) to support local/remote CUDA kernel launches. In both
scenarios, the GPU faces multiple pending kernel launch requests. Kernelet can be applied to schedule those pending kernels.

Our study mainly considers the throughput issues of
sharing a single GPU. Kernelet can be extended to multiple GPUs with
a workload dispatcher to \jianlong{each individual GPU}.

We have made the following
assumptions on the kernels.
\begin{enumerate}
  \item [1.] We target at the concurrent kernel
  executions on the shared GPU. The kernels are usually throughput
  oriented, with flexibility on the response time for scheduling.
  Still, we do not assume the a priori knowledge of the order of the
  kernel arrival.
  \item [2.] Thread blocks in a kernel are independent with each other.
  This assumption is mostly true for the GPGPU kernels due to SPMD programming model. Most kernels in NVIDIA SDK and benchmarks like
  Parboil~\cite{impact2007parboil} do not have dependency among thread blocks in the same
  kernel. This assumption ensures that our slicing technique on a given kernel is
  safe. The data dependency among thread
  blocks can be identified with standard static or dynamic program
  analysis.
\end{enumerate}

We formally define the terminology in Kernelet.

{\bf Kernel.} A kernel $\mathcal{K}$ consists of $k$ thread blocks
with IDs, 0, 1, 2, ..., $(k-1)$.

{\bf Slice.} A slice is a subset of the thread blocks of a launched
kernel. Block IDs of a
slice is continuous in the grid index space. The size of a slice $s$
is defined as the number of thread blocks contained in the slice.

{\bf Slicing plan.} Given a kernel $\mathcal{K}$, a slicing plan
$\mathcal{S} (\mathcal{K})$ is a scheme slicing $\mathcal{K}$ into a
sequence of $n$ slices ($\textfrc{s}_0$, $\textfrc{s}_1$,
$\textfrc{s}_2$, ..., $\textfrc{s}_{n-1}$). We denote the slicing plan
to be $\mathcal{K}$=$\textfrc{s}_0$, $\textfrc{s}_1$,
$\textfrc{s}_2$, ..., $\textfrc{s}_{n-1}$.

{\bf Co-schedule.} Co-schedule $\textfrc{cs}$ defines concurrent execution of
$n$ ($n \geq 1$) slices, \jianlong{denoted as $\textfrc{s}_0$, ...,
$\textfrc{s}_{n-1}$. All the $n$ slices are active on the GPU.}

{\bf Scheduling plan.} Given a set of $n$ kernels $\mathcal{K}_0$,
$\mathcal{K}_1$, ..., $\mathcal{K}_n$, a scheduling plan
$\mathcal{C}$ ($\textfrc{cs}_0$, $\textfrc{cs}_1$, ...,
$\textfrc{cs}_{n-1}$) determines a sequence of co-schedules in their
order of execution. $\textfrc{cs}_i$ is launched before
$\textfrc{cs}_j$ if $i<j$. All thread blocks of the $n$ kernels
occur in one of the co-schedules once and only once. A scheduling
plan embodies a slicing plan for each kernel.

We define the performance benefit of co-scheduling $n$ kernels to be
the co-scheduling profit ($\mathit{CP}$) in Eq.~(\ref{eq:cp}).
$IPC_{i}$ and $cIPC_{i}$ are IPC (Instruction Per Cycle) for
sequential execution and concurrent execution of kernel $i$
respectively. Our definition is similar to those in the previous
studies on CPU multi-threaded co-scheduling~\cite{jiang2008analysis,
SMTSymbiotic}.
\begin{equation}\label{eq:cp}
\mathit{CP}=1-\frac{1}{\sum\limits_{i=1}^{n}\frac{cIPC_{i}}{IPC_{i}}}
\end{equation}

{\bf Problem definition.} Given a set of kernels for execution, the
problem is to determine the optimal scheduling plan (and slicing) so
that the total execution time for those kernels is minimized. That
corresponds to the maximized throughput. Given a set of $n$ kernels
$\mathcal{K}_0$, $\mathcal{K}_1$, ..., $\mathcal{K}_{n-1}$, we aim
at finding the optimal scheduling plan $\mathcal{C}$ for a minimized
total execution time of
$\mathcal{C}($$\mathcal{S}_0(\mathcal{K}_0)$,
$\mathcal{S}_1(\mathcal{K}_1)$, ...,
$\mathcal{S}_{n-1}(\mathcal{K}_{n-1})$), or maximized co-scheduling
profit. Note, in the shared GPU environment, the arrival of new
kernels trigger the recalculation of the optimization on the kernel
residuals and new kernels.

\section{System Overview}\label{sec:systemdesign} In this section, we
present the rationales on the Kernelet design, followed by an
overview of the Kernelet runtime.

\subsection{Design Rationale}
Since kernels are submitted in an ad-hoc manner in our application
scenarios, the scheduling decision has to be made at real time. The
optimization process should take the newly arrived kernels into
consideration. Moreover, our runtime system of slicing and
scheduling should be designed with light overhead. That is, the
overhead of slicing and scheduling should be small compared with
their performance gain.

Unfortunately, finding the optimal slicing plans and scheduling plan
is a challenging task. The solution space for such candidate plans
is large. For slicing a kernel, we have the factors including the
number of slices as well as the slice size. For scheduling a set of
slices, we can generate different scheduling plan with co-scheduling
slices from different kernels. All those factors are added up into a
large solution space. Considering newly arrival kernels makes the
huge scheduling space even larger.

Due to the real-time decision making and light-weight requirements,
it is impossible to search the entire space to get a globally
optimal solution. The classic Monte Carlo simulation methods are not
feasible because they usually exceed our budget on the runtime
overhead and violate real-time decision making requirement. There
must be a more elegant compromise between the optimality and runtime
efficiency. Our solution to the scheduling problem is detailed in
Section~\ref{sec:SchedulingPlan}.

Given the complexity of dynamic slicing and scheduling in concurrent
kernel executions, we make the following considerations.

First, the scheduling considers two kernels only. Previous
studies~\cite{jiang2008analysis} on the CPU have shown that when
there are more than two co-running jobs, finding the optimal
scheduling plan becomes an NP-complete problem even ignoring the job
length differences and rescheduling. Following previous
studies~\cite{jiang2008analysis, gpu_consolidation_hpdc11}, we make our
scheduling decision co-scheduling two kernels only.

Second, once we choose two kernels to schedule, their slice sizes
keep unchanged until either kernel finishes. The suitable slice size
is determined according to our performance model~\ref{subsec:model}.

\subsection{System Overview}
We develop Kernelet as a runtime system to generate the slicing plan
and scheduling plans for the optimized throughput.

\begin{figure}
  \centering
  \includegraphics[width=0.48\textwidth]{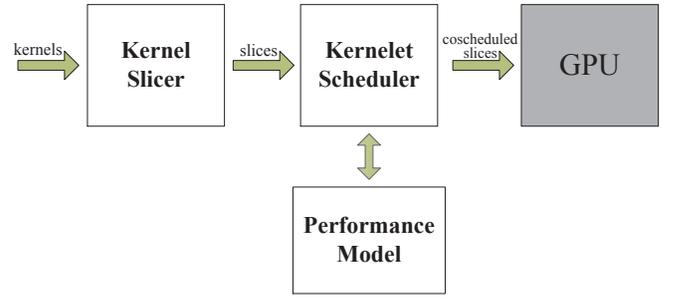}
  \caption{Design overview of Kernelet.}\label{fig:kerneletOverview}
\end{figure}

Figure~\ref{fig:kerneletOverview} shows an overview of Kernelet.
Kernels are submitted and are temporarily buffered in a kernel queue
for further scheduling. Usually, a kernel is submitted in the form
of binary (SASS) or PTX code. Submitted kernels are first
preprocessed by kernel slicer to \jianlong{determine the smallest
slice size for a given overhead limit}. If the kernel has been
submitted before, we simply use the smallest slice size in the
previous execution. Kernelet's scheduler determines the scheduling
plan based on the results of performance model, which estimates the
performance of slices from two different kernels in a probabilistic
manner. Once the scheduling plan is obtained, slices are dispatched
for execution on the GPU. We will describe the detailed design and
implementation of each component in the next section.

\section{Kernelet Methodology}\label{sec:methodology}
In this section, we first describe our kernel slicing mechanism. Next, we
present our greedy scheduling algorithm, followed by our pruning
techniques on the co-schedule space and the description of the
performance model.

\subsection{Kernel Slicing}\label{sec:SlicingMechanism}
The purpose of kernel slicing is to divide a kernel into multiple
slices so that the finer granularity of each slice as a kernel in
the scheduling can create more opportunities for time sharing.
Moreover, we need to determine the suitable slice size for
minimizing the slicing overhead (i.e., between the total execution
time of all slices and the kernel execution time) . Particularly, we
experimentally determine the suitable slice size to be the minimum
slice so that the overhead is not greater than $p\%$ of the kernel
execution time. In this study, $p\%$ is set to be \jianlong{2}\% by
default. We focus on the implementation of slicing on PTX or SASS
code. Note that, warps within the same thread block usually have
data dependency with each other, e.g., with the usage of shared
memory. That is why we choose thread blocks as the units for
slicing, instead of warps.

\begin{figure}
        \centering
        \begin{subfigure}{0.240\textwidth}
                \centering
                \includegraphics[width=\textwidth]{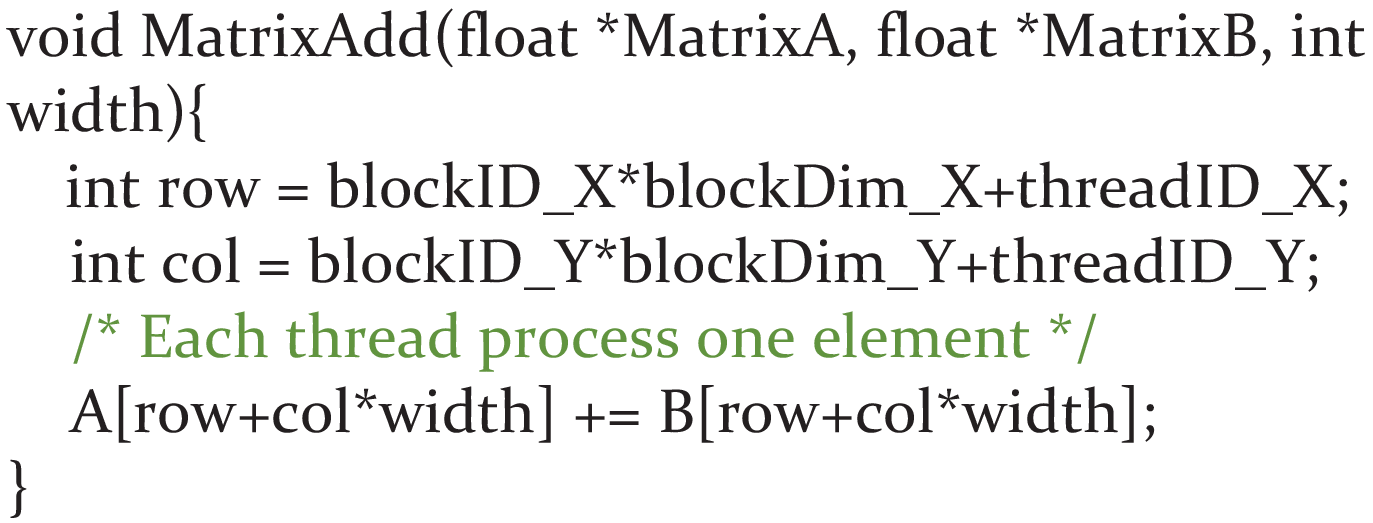}
                \vspace{-2ex}
                \caption{Each thread access the
corresponding matrix element using block and thread indices.}
                \label{fig:slicingA}
        \end{subfigure}
        \begin{subfigure}{0.240\textwidth}
                \centering
                \includegraphics[width=\textwidth]{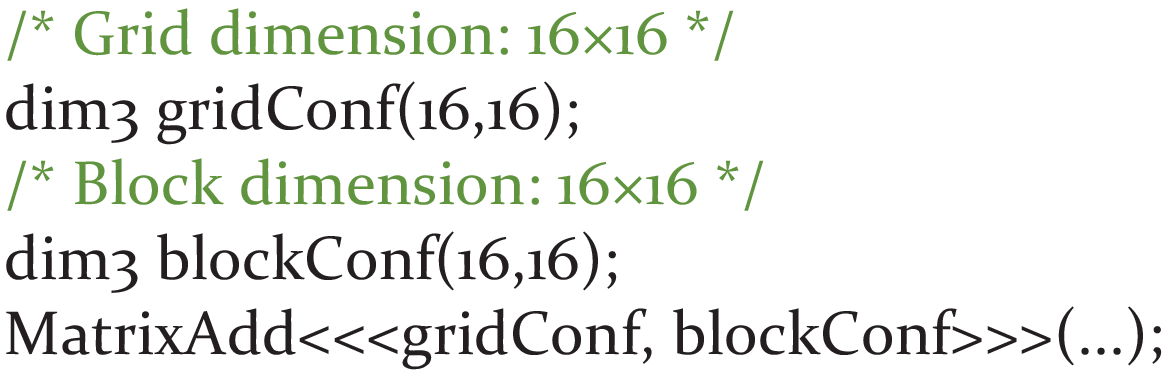}
                \vspace{-2ex}
                \caption{Launch the same number
threads as the number of matrix elements.}
                \label{fig:slicingB}
        \end{subfigure}\\
        \begin{subfigure}{0.240\textwidth}
                \centering
                \includegraphics[width=\textwidth]{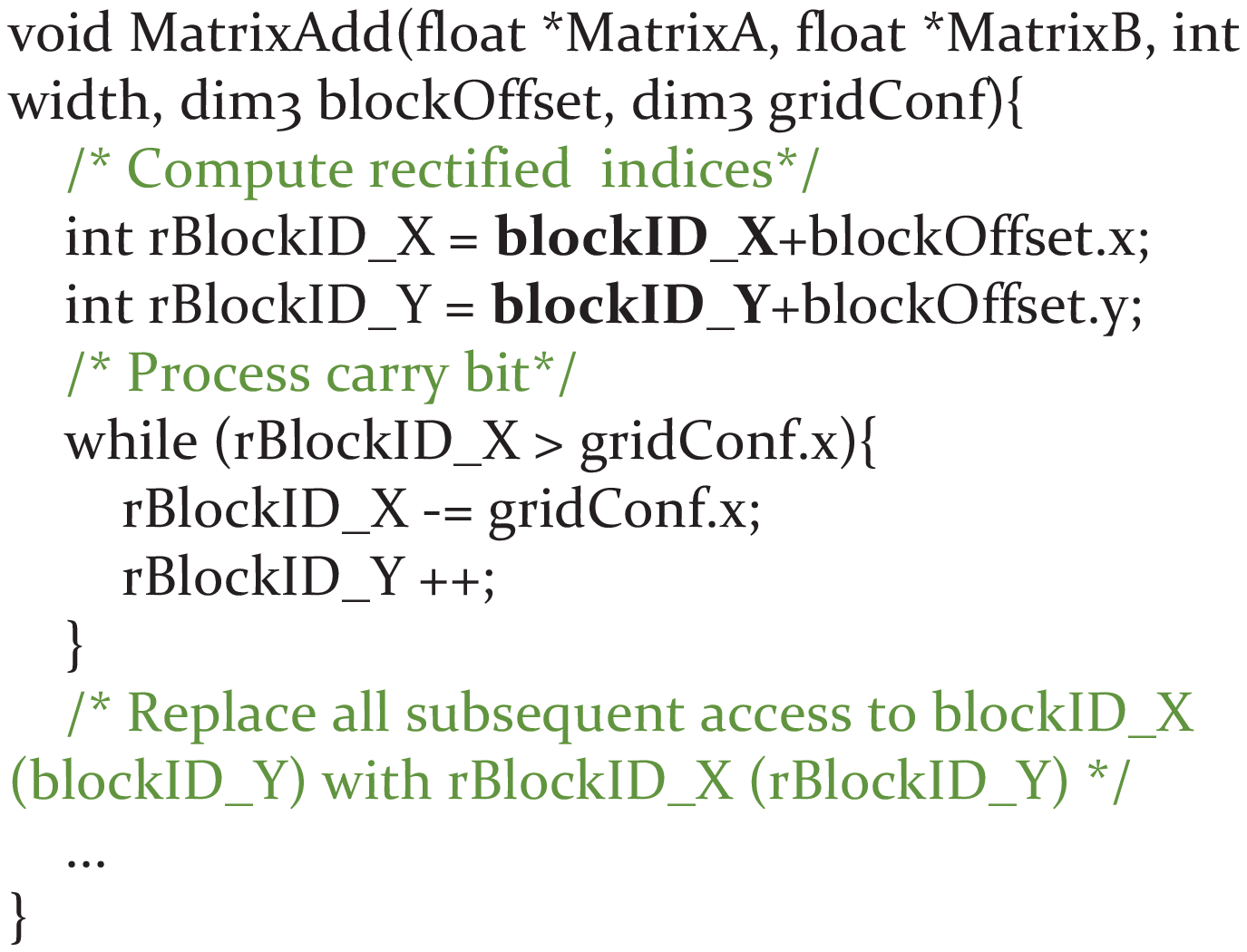}
                \vspace{-2ex}
                \caption{Sliced kernel with rectified thread block indices.}
                \label{fig:slicingC}
        \end{subfigure}
        \begin{subfigure}{0.240\textwidth}
                \centering
                \includegraphics[width=\textwidth]{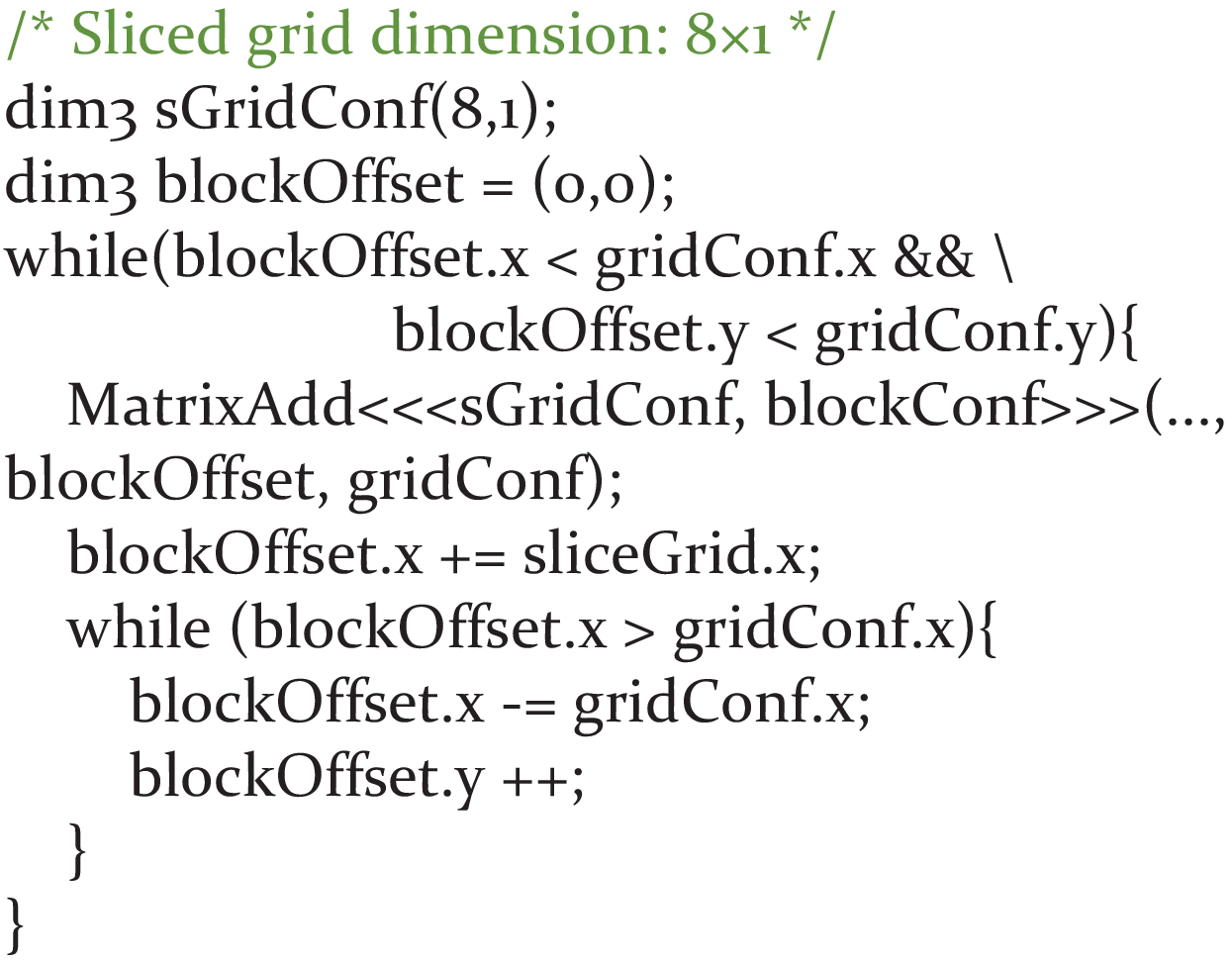}
                \vspace{-2ex}
                \caption{Launch all slices of a
kernel in a loop.}
                \label{fig:slicingD}
        \end{subfigure}

        \caption{An example of kernel slicing.}\label{fig:slicing}
\end{figure}

Figure~\ref{fig:slicing} illustrates the sliced execution of
$\mathit{MatrixAdd}$ with pseudo code. In the example,
$\mathit{MatrixAdd}$ is \jianlong{a GPU kernel} to add two
$256\times256$ matrices \jianlong{and each thread adds one element
pair from the input matrices.} Based on the matrix size,
$\mathit{MatrixAdd}$ is configured to launch with $16\times 16$
thread blocks in total and each block has $16\times16$ threads.
\jianlong{Figures~\ref{fig:slicingA} and~\ref{fig:slicingB}
illustrate the definition and launch of the original kernel,
respectively.} In comparison, the sliced version of the kernel
launches a slice with 8 thread blocks each time. \jianlong{The
built-in thread block indices (denoted as $\mathit{blockID\_X}$ and
$\mathit{blockID\_Y}$) of the sliced kernel are in a smaller index
space ($\{(x,0) | 0 \leq x < 8 \}$) compared with the original index
space ($\{(x,y) | 0 \leq x < 16\ and\ 0 \leq y < 16\}$). To make the
slices execute as individual kernels, we apply a procedure called
\emph{index rectification}. As shown in Figure~\ref{fig:slicingC},
we add an offset value to the thread block indices and obtain the
rectified index values. The rectified indices are used to replace
all subsequent accesses to the built-in indices. On the CPU side, we
launch the slices in a loop and adjust the offset values for each
slice launch (Figure~\ref{fig:slicingD}).}

Kernelet automatically implements slice rectification without any
user intervention. With PTX or SASS code as input, Kernelet does not
require source code. Kernelet interprets and modifies the PTX/SASS
code at runtime. The resulting PTX code is compiled to GPU
executables by the GPU driver, and the SASS code is assembled using
the open source Fermi assembler Asfermi~\cite{asfermi}. Kernelet
stores the rectified block index in registers, and replaces all
references to those built-in variables with the new registers. Since
this process may use more registers. Kernelet tries to minimize the
register usage by adopting the classic register minimization
techniques~\cite{poletto1999linear,chaitin1982register}, e.g.,
variable liveness analysis. With register optimizations, register
usage by slicing keeps unchanged in most of our test cases in
experiments. \jianlong{Note, kernel slicing only requires a single
scan on the input code and the runtime overhead is negligible.}

\subsection{Scheduling}\label{sec:SchedulingPlan}

According to our design rationales, our scheduling decision is made
on the basis of two kernels, to avoid the complexity of scheduling three or more kernels
as a whole; the slice sizes of the two kernels are tuned for GPU
utilization so that their execution times in the concurrent kernel
execution are close. Thus, we develop a greedy scheduling
algorithm, as shown in Algorithm~\ref{alg:overall}. The scheduling
algorithm considers new arrival kernels in Lines 2--3 in the main
algorithm. The main procedure calls the procedure
\emph{FindCoSchedule} to obtain the optimal co-schedule in Line 5.
The co-schedule is represented in four parameters $<$
$\mathcal{K}_1$, $\mathcal{K}_2$, $\mathit{size}_1$,
$\mathit{size}_2$ $>$, where $\mathcal{K}_1$ and $\mathcal{K}_2$
denotes the two selected kernels; $\mathit{size}_1$ and
$\mathit{size}_2$ represents the slice sizes accordingly. We use the
same co-schedule if the kernels pending for execution do not change,
or both kernels still have thread blocks.

\begin{algorithm}
\caption{Scheduling algorithm of Kernelet}
\begin{algorithmic}[1]\label{alg:overall}
    \STATE Denote $\mathcal{R}$ to be the set of kernels pending for
    executions;
    \IF{A new kernel $\mathcal{K}$ comes}
    \STATE Add $\mathcal{K}$ into $\mathcal{R}$;
    \ENDIF
    \WHILE{$\mathcal{R}$!=null}
    \STATE $<$ $\mathcal{K}_1$, $\mathcal{K}_2$, $\mathit{size}_1$, $\mathit{size}_2$ $>$=\emph{FindCoSchedule}($\mathcal{R}$);
    \STATE Denote the co-schedule to be $c$;
    \STATE Execute $c$ on the GPU;
    \WHILE{$\mathcal{R}$ does not change, or $\mathcal{K}_1$ and $\mathcal{K}_2$ both still have thread blocks}
    \STATE Generate co-schedule according to $c$ and execute it on the GPU;
    \ENDWHILE
    \ENDWHILE
\end{algorithmic}
{\bf Proc. }\emph{FindCoSchedule($\mathcal{R}$)}\\ {\bf Function:
generate the optimal co-schedule from $\mathcal{R}$.}
\begin{algorithmic}[1]
    \STATE Generate the candidate space for co-schedules $\mathcal{C}$;
    \STATE Perform pruning on $\mathcal{C}$ according to the
    computation and memory characteristics of input kernels;
    \STATE Apply the performance model (Section~\ref{subsec:model}) to compute $\mathit{CP}$ for all the co-schedule in
    $\mathcal{C}$;
    \STATE Obtain the optimal co-schedule with the maximized
    $\mathit{CP}$;
    \STATE Return the result co-schedule;
\end{algorithmic}
\end{algorithm}

In Procedure \emph{FindCoSchedule}, we first consider the entire
candidate space consisting of co-schedules on pair-wise kernel
combinations. Because the space may consider of $\frac{N(N-1)}{2}$ co-schedules
($N$ is the number of kernels for consideration), it is desirable to
reduce the search space.
Therefore, we perform pruning according to the computation and
memory characteristics of input kernels. We present the details of
pruning mechanisms in Section~\ref{subsec:pruning}. After pruning,
we apply the performance model (Section~\ref{subsec:model}) to
estimate the $\mathit{CP}$ for all co-schedules, and pick the one
with the maximized $\mathit{CP}$ for executing on the GPU.

\subsection{Co-scheduling Space Pruning}\label{subsec:pruning}

Given a set of co-schedules as input, we aim at developing pruning
techniques to remove the co-schedules that are not ``promising" to
deliver performance gain. So that the overhead of running the performance model is avoided. The basic idea to identify the key
performance factor of a single kernel that affects the throughput of concurrent kernel
executions on the GPU.

There are many factors affecting the GPU performance. According to
the CUDA profiler, there are around 100 profiler counters and
statistics for the performance tuning. For an effective scheduling
algorithm, there is no way of considering all these factors.
Following a previous work~\cite{SMTModelContentionICCD05}, we use
the regression model to explore the correlation between the above
mentioned factors and $\mathit{CP}$. The input values are obtained
from single kernel executions. Through detailed performance studies,
we find that instruction throughput and memory bandwidth utilization
are the most correlated performance factors with co-scheduling
friendliness. We define PUR (Pipeline Utilization Ratio) and MUR
(Memory-bandwidth Utilization Ratio) to characterize user submitted
kernels. High PUR means the instruction pipeline is highly utilized
and there is \jianlong{little room} for performance improvement.
High MUR means a large number of memory requests are being processed
and memory latency is high. PUR and MUR is calculated as follows,
\begin{equation*}
PUR=\frac{Instruction\_Executed}{Time\times Frequency\times
Peak\_IPC}
\end{equation*}

\begin{equation*}
MUR=\frac{Dram\_Reads+Dram\_Writes}{Time\times Frequency\times
Peak\_MPC}
\end{equation*}

$Peak\_IPC$ and $Peak\_MPC$ represent peak number of instructions
and memory requests per cycle respectively. $Instruction\_Executed$ is the total
number of instructions executed. $Dram\_Reads$ and $Dram\_Writes$ are the numbers of read and write
requests to DRAM respectively.

\begin{figure}\label{fig:application_scenario}
        \centering
        \begin{subfigure}{0.24\textwidth}
                \centering
                \includegraphics[width=\textwidth]{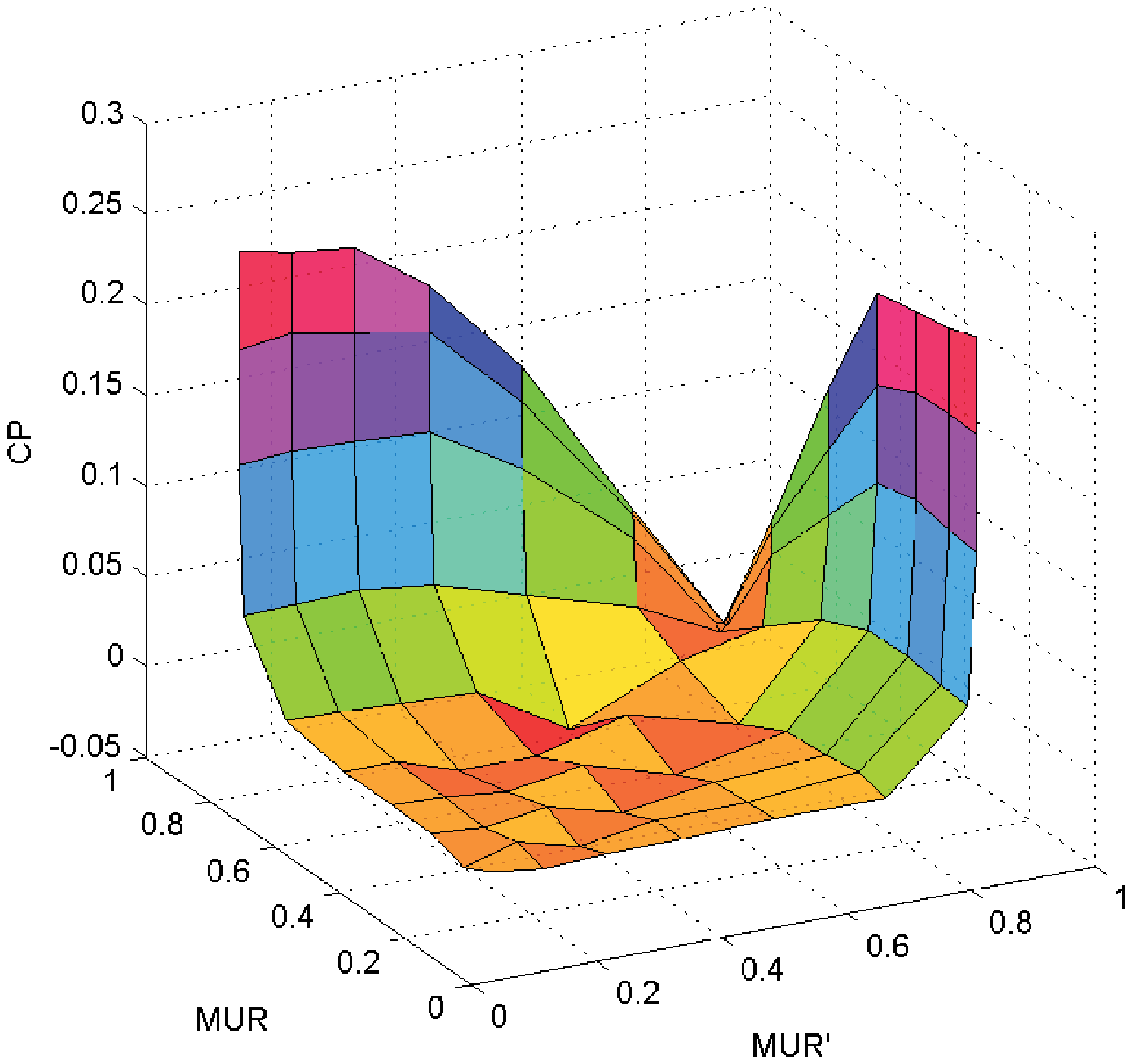}
                \caption{MUR and $\mathit{CP}$}
                \label{fig:mur_speedup}
        \end{subfigure}
        \begin{subfigure}{0.24\textwidth}
                \centering
                \includegraphics[width=\textwidth]{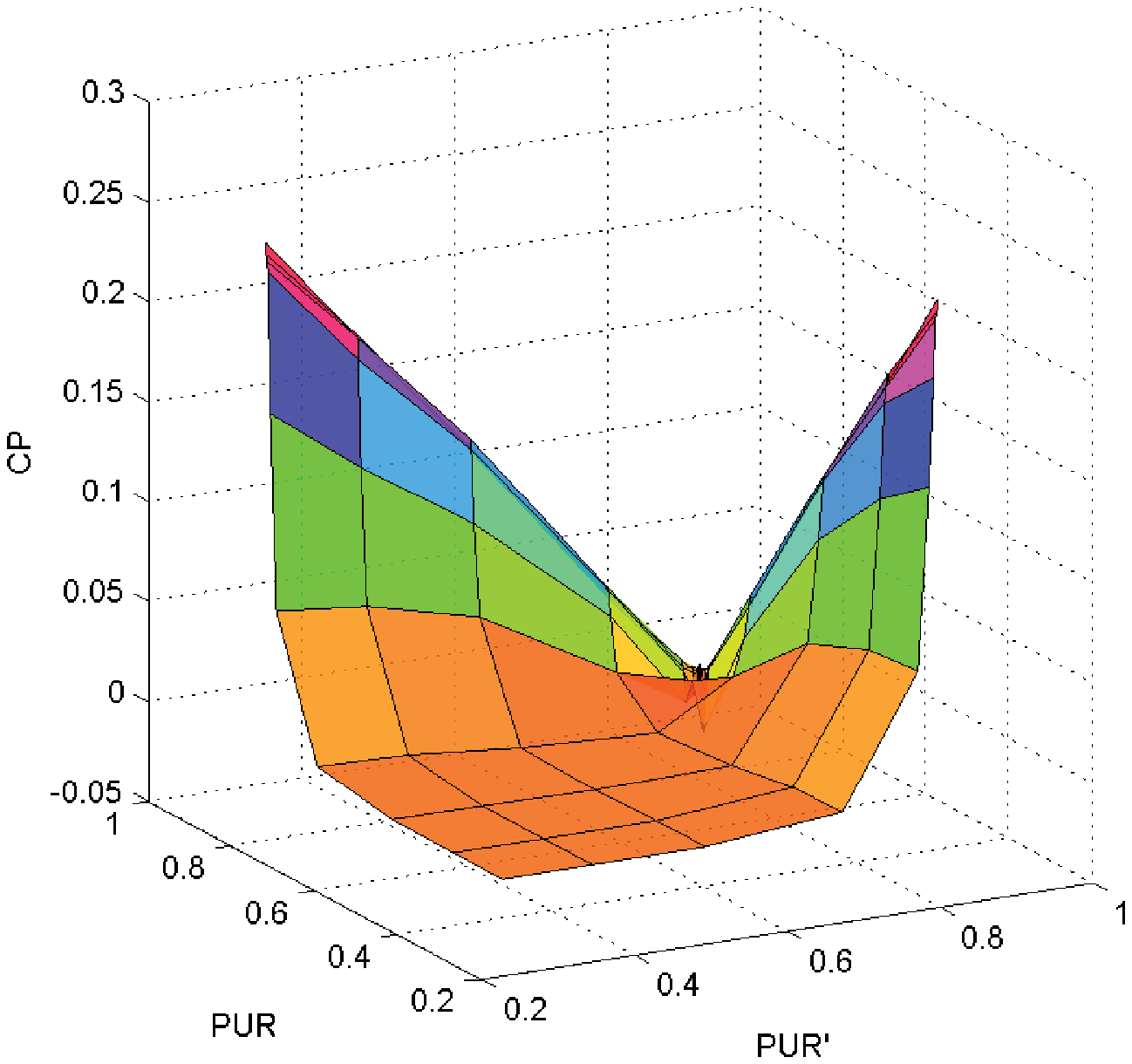}
                \caption{PUR and $\mathit{CP}$}
                \label{fig:pur_speedup}
        \end{subfigure}
        \caption{Correlation between MUR/PUR and $\mathit{CP}$}\label{fig:pur_mur}
\end{figure}

We build a set of testing kernels to demonstrate the correlation between PUR/MUR and $\mathit{CP}$.
A testing kernel is a mixture of memory and computation instructions.
We tune the respective instruction ratios to obtain kernels with different memory and computational characteristics.
The single kernel execution PURs and MURs of the testing kernels
are in the range of $[0.26, 0.83]$ and $[0.07, 0.84]$ respectively.
Figure~\ref{fig:pur_mur} shows the strong correlation between
MUR/PUR and $\mathit{CP}$. The observation conforms to our expectation.
First, if one kernel has high PUR while the other kernel has low
PUR, the former kernel is able to utilize the idle cycles exposed by
the latter kernel when co-scheduled. Second, co-scheduling kernels
with complementary memory requirements (one kernel has low MUR and
the other kernel has high MUR) will alleviate memory contention and
reduce idle cycles exposed by long latency memory operations.

\jianlong{In summary, our pruning rule is to remove the co-schedules
where the two kernels have close PUR or MUR values. We set two
threshold values $\alpha_p$ and $\alpha_m$ for PUR and MUR,
respectively. That means, we prune the co-schedule if the two
kernels have PUR difference lower than $\alpha_p$, or have MUR
difference lower than $\alpha_m$. Note, if all the co-schedules are
pruned, we need to increase $\alpha_p$ or $\alpha_m$. We
experimentally evaluate the impact of those two threshold values in
Section~\ref{sec:evaluation}.}

\subsection{Performance Model}\label{subsec:model}
We need a performance model for two purposes: firstly, to select the
two kernels for co-schedule; secondly, to determine the number of
thread blocks for each kernel \jianlong{in the co-schedule} (i.e.,
the slice size). Previous performance models on the
GPU~\cite{kimmodel_isca09, kimmodel_ppopp12,owensModel} assume a
single kernel on the GPU, and are not applicable to concurrent
kernel executions. They generally assume that the thread blocks
execute the same instruction in a round-robin manner on an SM.
However, this is no longer true on concurrent kernel executions. The
thread blocks from different kernels have interleaving executions,
which cause non-determinism on the instruction execution flow. It is
not feasible to statically predict the interleaving warp executions
for multi-kernel executions. To capture the non-determinism, we
develop a probabilistic performance model to estimate the
performance of co-schedule. Our cost model has very low runtime
overhead, because it uses a series of simple parameters as input and
leverages the Markov chain theory to get the performance of
concurrent kernel executions.

Table~\ref{tb:para} summarizes the notations used for our
performance model.

\begin{table}
\centering
   \caption{Parameters and notations in the performance model\label{tb:para}}{
  \begin{tabular}{|c|p{6cm}|}
    \hline
    Para.     & Description \\\hline
    $W$         & Maximum number of active warps    \\\hline
    $Round$     & A warp scheduling cycle that all ready warps are
    served by the warp scheduler \\\hline
    $R_{m}  $ & Memory instruction ratio     \\\hline
    $P_{r\rightarrow r}$ & Probability that a ready warp remains ready\\\hline
    $P_{r\rightarrow i}$ & Probability that a ready warp transits to idle\\\hline
    $P_{i\rightarrow r}$ & Probability that an idle warp transits to ready\\\hline
    $P_{i\rightarrow i}$ & Probability that an idle warp remains in idle\\\hline
    $N_{r\rightarrow r}$ & Number of ready warps that remain ready\\\hline
    $N_{r\rightarrow i}$ & Number of ready warps that transit to idle\\\hline
    $N_{i\rightarrow r}$ & Number of idle warps that transit to ready\\\hline
    $N_{i\rightarrow i}$ & Number of idle warps that remain idle\\\hline
    $L$       & Average memory latency (cycle)\\\hline
    $B$       & GPU global memory bandwidth (requests/cycle)\\\hline
    $S_i$     & $S_i$ corresponds the state where $i$ warps are idle on the SM ($i=0, 1, \ldots , W$)\\\hline
    $P_{ij}$  & $P$ is the Markov chain transit matrix. Entry $P_{ij}$ of $P$ represents the probability of transiting from state $S_i$ to state $S_j$. \\\hline
  \end{tabular}
  }
\end{table}

Since the GPU adopts SPMD model, we use the performance estimation
of one SM to represent the aggregate performance of all SMs on the
GPU. We model the process of kernel instruction issuing as a stochastic process and
devise a set of states for an SM during execution. By modeling the
SM state, we first develop our Markov chain based models for
single-kernel executions (homogeneous workloads), and then extend it
to concurrent kernel executions (heterogeneous workloads).

For presentation clarity, we begin with our description on the model
with the following assumptions, and relax those assumptions at the
end of this section. First, we assume that all the memory requests
are coalesced. This is the best case for memory performance. We will
relax this assumption by considering both coalesced and uncoalesced
memory accesses. Second, we assume that the GPU has a single warp
scheduler. We will extend it to the GPU with multiple warp
schedulers.

{\bf Homogeneous Workloads.} We first investigate the performance of
a single kernel executed on the GPU and each SM accommodates $W$
active warps at most.

\begin{figure}
  \centering
  \includegraphics[width=0.50\textwidth]{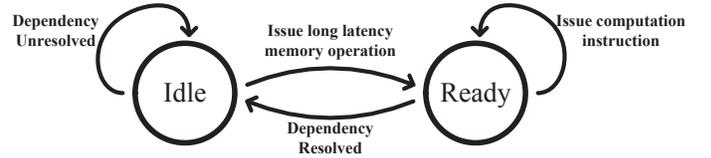}
  \caption{Warp state transition diagram.}
  \label{fig:warpTransition}
\end{figure}

A warp can be in two states: idle or ready. An idle warp is stalled
by memory accesses, and a ready warp has one instruction ready for
execution. Its transition is illustrated in
Figure~\ref{fig:warpTransition}. When a warp is currently in the
ready state, we have two cases for state transitions by definition:
\begin{itemize}
  \item remaining in the ready state with the probability of $P_{r\rightarrow r}=1-R_{m}$.
  \item transiting to the idle state with the probability of $P_{r\rightarrow i}=R_{m}$.
\end{itemize}

When a warp is currently in the idle state, we also have two cases
for state transitions:
\begin{itemize}
  \item  transiting to the ready state with the probability of $P_{i\rightarrow
  r}=\frac{1}{\frac{L}{W-I}}=\frac{W-I}{L}$, where $I$ is the number
  of idle warps on the SM.
  \item  remaining in the idle state with the probability of $P_{i\rightarrow i}=1-P_{i\rightarrow r}$.
\end{itemize}

We specifically define the time step and state transition of the
Markov chain model to capture the GPU architectural features. GPU
adopts a round-robin style scheduling~\cite{kimmodel_isca09}. In
each \emph{round}, the warp scheduler polls each warp to issue its
ready instructions so all ready warps can make progress. We model
the SM state with the number of idle warps. We denote $S_i$ to be
the SM state where $i$ warps are idle on the SM ($i=0, 1, \ldots ,
W$). Thus, we consider the state change of the SM in one round and
use round as time step in our Markov chain model. In each round,
every ready warp has an equal chance to issue instructions. In
contrast, models for the CPU assume that the CPU will keep executing
one thread until this thread is suspended.


We use $\mathit{IPC}$ to represent the throughput of the SM. Thus,
the number of idle warps on the SM is a key parameter for
$\mathit{IPC}$. Thus, we define the state of SM as the number of
idle warps on the SM (i.e., the state is $S_i$ when the number of
idle warps is $i$). More outstanding memory requests usually lead to
higher latency because of memory
contention~\cite{saramodel_ppopp12}. We adopt a linear memory model
to account for the memory contention effects. We calculate $L$ as
$L=L_{0}+\frac{B}{a_0\cdot S_i}+b_0$, where $a_0$ and $b_0$ are the
constant parameters in the linear model. We obtain $L_{0}$ and $B$
according to hardware specifications.

For homogeneous workload, the probabilities of state transitions are
the same for all ready warps in a round. We assume when SM transits
from $S_i$ to $S_j$, $N_{i\rightarrow r}$ idle warps transit to the
ready state and $N_{r\rightarrow i}$ ready warps transit to idle
state. The following conditions hold by definition.
\begin{equation}
\begin{cases}
0 \le N_{i\rightarrow r} \le S_i\\
0 \le N_{r\rightarrow i} \le W-S_i\\
N_{r\rightarrow i}-N_{i\rightarrow r}=S_j-S_i
\end{cases}
\end{equation}

With those constraints, there are multiple possible transitions to
transit from $S_i$ to $S_{j}$. Since the possible transitions are
mutually exclusive events, the probability of each state transition
$P_{ij}$ is calculated as the sum of the probabilities of all
possible transitions. With all entries of the transition matrix $P$
obtained, we can calculate the steady-state vector of the Markov
chain. This is done by finding the eigenvector $\pi$ corresponding
to the eigenvalue one for matrix $P$~\cite{papoulis1991stochastic}.
\begin{equation}\label{eq:steadyStateVector}
\pi=(\gamma_{0},\gamma_{1},..., \gamma_{W})
\end{equation}

In Equation~\ref{eq:steadyStateVector}, $\gamma_{i}$ is the
probabilities that the SM stays in state $S_i$ in each round, i.e.,
the probability there are $i$ idle warps in one round. The duration
of the time step is $(W-i)$ cycles since each of the $(W-i)$ ready
warps issue one instruction within the round. In the case $i=W$, the
round duration is one, indicating no warp is ready and the SM
experiences an idle cycle. Hence, the estimated $\mathit{IPC}$ is
the ratio of non-idle cycles given in equation~\ref{eq:modelIPC},
where $\sum_{i=0}^{W-1}\gamma_{i}(W-i)$ is the total non-idle cycles
and $\gamma_{W}$ is the total idle cycles.
\begin{equation}\label{eq:modelIPC}
IPC_{\mathcal{K}}=\frac{\sum_{i=0}^{W-1}\gamma_{i}(W-i)}{\sum_{i=0}^{W-1}
\gamma_{i}(W-i)+\gamma_{W} }
\end{equation}


{\bf Heterogeneous Workloads.} When there are multiple kernels
running concurrently, the model needs to keep track of the state of
each workload. Although we only consider two concurrent kernels
($\mathcal{K}_1$ and $\mathcal{K}_2$) in scheduling, our model can
be used to handle more than two kernels.

Suppose there are two kernels $\mathcal{K}_{1}$ and
$\mathcal{K}_{2}$, and $\mathcal{K}_{1}$ has $w_1$ active warps and
$\mathcal{K}_{2}$ has $w_2$ active warps $(w_1+w_2=W)$. The number
of possible states of the SM will be $(w_1+1)\times(w_2+1)$. The
state space is represented as a pair $(p,q)$ with $0 \le p \le w_1$
and $0\le q \le w_2$, where $p$ and $q$ are the numbers of idle
warps of $\mathcal{K}_{1}$ and $\mathcal{K}_{2}$ respectively. We
can calculate the probability of transiting from state
$(p_{i},q_{i'})$ to state $(p_{j},q_{j'})$ by first considering
individual workload state transition probability using the single
kernel model, and then calculating the SM state transition
probability. The state transitions of different kernels are
independent with each other, because the kernels are independent.
Then the SM state transition probability is the product of the
individual transition probabilities.

With Markov chain approach, we obtain the steady state vector
$\pi=\{\gamma_{(0,0)},\gamma_{(0,1)},...,\gamma_{(w_1,w_2)}\}$.
Next, we can obtain the $\mathit{IPC}$ of each workload using the
same method as the model in single-kernel executions,
\jianlong{except the parameters are defined and calculated in the
context of two kernels. For example, the round duration is equal to
the total number of ready warps of both kernels}. Individual IPCs of
$\mathcal{K}_{1}$ and $\mathcal{K}_{2}$ is calculated as the ratio
of non-idle cycles for each workload, as shown in
Eq.~(\ref{eq:ipc_k1}) and~(\ref{eq:ipc_k2}), respectively. The
concurrent IPC is the sum of individual IPCs Eq.~(\ref{eq:ipc}).
Then $\mathit{CP}$ can be obtained using Eq.~(\ref{eq:cp}).

\begin{equation}\label{eq:ipc_k1}
IPC_{\mathcal{K}_{1}}=\frac{\sum_{i=0}^{w_1-1}\sum_{i'=0}^{w_2}
\gamma_{(i,i') }\times
(w_1-i)}{\sum_{i=0}^{w_1}\sum_{j=0}^{w_2}\gamma_{(i,j)}\times
R_{(i,j)}}
\end{equation}

\begin{equation}\label{eq:ipc_k2}
IPC_{\mathcal{K}_{2}}=\frac{\sum_{i=0}^{w_1}\sum_{i'=0}^{w_2-1}
\gamma_{(i,i') }\times (w_2-i')}{\sum_{i=0}^{w_1}\sum_{i'=0}^{w_2}
\gamma_{(i,i')}\times R_{(i,i')}}
\end{equation}

\begin{equation}\label{eq:ipc}
C=IPC_{\mathcal{K}_{1}}+IPC_{\mathcal{K}_{2}}
\end{equation}

With the estimated IPC and $\mathit{CP}$, we now discuss how to
estimate the \jianlong{optimal} slice size ratio for two kernels. We
define the slice ratio which minimizes the execution time difference
of co-scheduled slices as the balanced slice ratio. By minimizing
the execution time difference, the kernel-level parallelism is
maximized. The execution time difference is calculated as $\Delta T$
in Eq.~(\ref{eq:executiontime}).
\begin{equation}\label{eq:executiontime}
\begin{array}{lcl}
\Delta T=\vert \frac{1}{\mathit{IPC_{\mathcal{K}_{1}}}}\times I_{\mathcal{K}_{1}}\times P_{\mathcal{K}_{1}} - 
\frac{1}{IPC_{\mathcal{K}_{2}}}\times I_{\mathcal{K}_{2}}\times P_{\mathcal{K}_{2}}\vert \\
\end{array}
\end{equation}
\noindent $I_{\mathcal{K}_{i}}$ and $P_{\mathcal{K}_{i}}$ represent
the number of instruction per block and the slice size of kernel
$\mathcal{K}_{i}$ ($i=1,2$) in number of thread blocks. Since
$P_{\mathcal{K}_{i}}$ is less than the maximal number of active
thread blocks, only a limited number of slice ratios need to be
evaluated to get the balanced
ratio. 

{\bf Uncoalesced Access.} So far, we assume that all memory accesses
are coalesced and each memory instruction results in the same number
of memory requests. However, due to the different address patterns,
memory instructions may result in a different amount of memory
requests. On Fermi GPUs, one memory instruction can generate 1 to 32
memory requests. Here we consider the two most common access
patterns: fully coalesced access, and fully uncoalesced access.
\jianlong{We extend our model to handle both coalesced and
uncoalesced accesses by defining} three states for a warp: ready,
stalled on coalesced access (uncoalesced idle), and stalled on
uncoalesced access (coalesced idle). The memory operation latency
depends on the memory access type. Since uncoalesced access
generates more memory traffic, its latency is higher than coalesced
access. We also use the linear model to estimate the latency. By
identifying the ratio of coalesced and uncoalesced memory
instructions, we can easily extend the two-state model to handle
three states and their state transitions. The Markov chain
performance model can be developed in a similar way. Distinguishing
between coalesced and uncoalesced accesses increases the accuracy of
our model.

{\bf Adaptation to GPUs with multiple warp schedulers.}
\jianlong{Our model assumes there is only one warp scheduler}.
New-generation GPUs can support more than one warp schedulers. The
latest Kepler GPU features four warp schedulers per SMX (SMX is the
Kepler terminology for SM)~\cite{kepler_white_paper}. We extend our
model to handle this case by deriving a single pipeline virtual SM
based on the parameters of the SMX. The virtual SM has one warp
scheduler, and its parameters such as active thread blocks and
memory bandwidth are obtained by dividing the corresponding
parameters of the SMX by the number of warp schedulers. This
virtual SM can still capture the memory and computation features of
a kernel running on the SMX. Experimental results in
Section~\ref{sec:evaluation} show that performance modeling on the
virtual SM provides a good estimation on the Kepler architectures.

There are two more issues that are worthwhile to discuss.

The first issue is on the efficiency of executing our model
\jianlong{at} runtime. We have developed mechanisms to make our
model more efficient without significantly sacrificing the model
accuracy. \jianlong{The $O(N^3)$ complexity of calculating the
steady state in Markov chain makes it hard to meet the online
requirement ($N$ is the dimension of the transition matrix).} To
reduce the computational complexity, we consider the thread block as
a scheduling unit, instead of considering individual warps. In this
way, the computational complexity is significantly reduced, and time
cost of our model is negligible to the GPU kernel execution time.

The second issue is on getting the input for the model. Our current
approach is based on hardware profiling of a small number of thread
blocks from a single kernel. Thus, the pre-execution is only a very
small part of the kernel execution. From profiling, we can obtain
the number of memory instructions issued and the total number of
instructions executed, and calculate $\mathit{Rm}$ as their ratio.

In summary, our probabilistic model has captured the inherent
non-determinism in concurrent kernel executions. First, it simply
requires only a small set of profiling inputs on the memory and
computation characteristics of individual kernels. Second, with
careful probabilistic modeling, we develop a performance model that
is sufficiently accurate to guide our scheduling decision. The
effectiveness of our model will be evaluated in the experiments
(Section~\ref{sec:evaluation}).

\section{Evaluation}\label{sec:evaluation}
In this section, we present the experimental results on evaluating Kernelet on latest GPU architectures.

\subsection{Experimental Setup}\label{subsec:experimentalSetup}
We have conducted experiments on a workstation equipped with
\jianlong{one NVIDIA Tesla C2050 GPU, one NVIDIA GTX680 GPU, two
Intel Xeon E5645 CPUs and 24GB RAM.} Table~\ref{tb:hardware} shows
some architectural features of C2050 and GTX680. We note that C2050
and GTX680 are based on Fermi and Kepler architectures,
respectively. One \jianlong{C2050} SM has two warp schedulers, and
each can serve half a warp per cycle (with a theoretical
$\mathit{IPC}$ of one). In contrast, one \jianlong{GTX680} SMX
features four warp schedulers and each warp scheduler can serve one
warp per cycle (with a theoretical $\mathit{IPC}$ of eight
considering its dual-issue capability). Our implementation is based
on GCC 4.6.2 and NVIDIA CUDA toolkit 4.2.

\begin{table}\footnotesize
\centering \caption{\label{tb:hardware}GPU configurations.}
\begin{tabular}
{|l|c|c|c|c}\hline
                       & C2050         & GTX680     \\\hline
Architecture           & Fermi GF110    & Kepler GK104 \\\hline
Number of SMs          & 14            & 8        \\\hline

Number of cores per SM & 32            & 192      \\\hline

Core frequency (MHz)   & 1147          & 706      \\\hline

Global memory size (MB)& 3072          & 2048     \\\hline

Global memory bandwidth (GB/s)& 144    & 192 \\\hline

\end{tabular}
\end{table}

{\bf Workloads.} We choose eight benchmark applications with
different memory and computation intensivenesses. Sources of the
benchmarks include the CUDA SDK, the Parboil
Benchmark~\cite{impact2007parboil}, the CUSP library~\cite{Cusp} and
our home grown applications. Table~\ref{tb:description} describes
the details of each application, including input settings and thread
configurations of
the most time-consuming kernel on C2050.

Table~\ref{tb:benchmarks} shows the memory and computation
characteristics of the most time-consuming kernel of each application on both C2050 and
GTX680. We observed
that the PUR/MUR values are stable as we vary the input sizes (as
long as the input size is sufficiently large to keep the GPU
occupancy high).

\begin{table}\scriptsize
\centering \caption{\label{tb:description}Specification of benchmark
applications and thread configuration (\#threads per thread block
$\times$ \#thread blocks).}
\begin{tabular}
{|p{1.5cm}|p{1.8cm}|p{2.1cm}|p{1.9cm}|}
  \hline
  Name                                        & Description                                                     & Input settings                                                            & Thread configuration on C2050   \\\hline
  Pointer Chasing (PC)                        & Traversing an array randomly                                    & Index values for 40 million accesses                                      & 256 $\times$ 16384    \\\hline
  Sum of Absolute Differences (SAD)           & An operation used in MPEG encoding                              & Image with $1920\times 1072$ pixels                                       & 32 $\times$ 8048 \\\hline
  Sparse Matrix Vector Multiplication (SPMV)  & Multiplying a sparse matrix with a dense vector.                & A 131072$\times$81200 matrix with 16 non-zero elements per row on average & 256 $\times$ 16384    \\\hline
  Stencil (ST)                                & Stencil operation on a regular 3-D grid                         & 3D grid with 134217728 points                                             & 128 $\times$ 16384 \\\hline
  Matrix Multiplication (MM)                  & Multiplying two dense matrices                                  & One $8192\times 2048$ matrix, the other $2048\times 2048$                 & 256 $\times$ 16384     \\\hline
  Magnetic Resonance Imaging - Q (MRIQ)       & A matrix operation in magnetic resonance imaging                & 2097152 elements                                                          & 256 $\times$ 8192\\\hline
  Black Scholes (BS)                          & Black-Scholes Option Pricing                                    & 40 million                                                                & 128 $\times$ 16384 \\\hline
  Tiny Encryption Algorithm (TEA)             & A thread block cipher                                           & 20971520 elements                                                         & 128 $\times$ 16384  \\\hline

\end{tabular}
\end{table}

\begin{table}\scriptsize
\centering \caption{\label{tb:benchmarks}Memory and computational
characteristics of benchmark applications.}
\begin{tabular}
{|l|p{0.6cm}|p{0.6cm}|p{1.1cm}|p{0.6cm}|p{0.6cm}|p{1.1cm}|}
  \hline
  \multirow{2}{*}{Benchmarks} & \multicolumn{3}{c|}{C2050}   & \multicolumn{3}{c|}{GTX680}\\\cline{2-7}
      & PUR & MUR & Occupancy & PUR & MUR & Occupancy \\\hline
  PC & 0.0096 & 0.1404 & 100\% & 0.0072 & 0.1746 & 100\% \\\hline
  SAD & 0.1498 & 0.1120 & 16.7\% & 0.1062 & 0.1351 & 25\% \\\hline
  SPMV & 0.3464 & 0.003 & 100\% & 0.3027  & 0.0043 & 100\% \\\hline
  ST & 0.3629 & 0.1156 & 66.7\% & 0.2016 & 0.1179 &100\% \\\hline
  MM& 0.5804 & 0.0161 & 67.7\% &0.5321  & 0.0569 & 100\% \\\hline
  MRIQ & 0.8539 & 0.0002 & 83.3\% & 1.6784  & 0.0007 & 100\% \\\hline
  BS &0.8642 & 0.0604 & 67.7\% &1.2007 & 0.1323 & 100\% \\\hline
  TEA & 0.9978 & 0.0196 & 67.7\% & 1.1417  & 0.0353 & 100\% \\\hline

\end{tabular}
\end{table}

To assess the impact of kernel scheduling under different mixes of
kernels, we create four groups of kernels namely CI, MI, MIX and ALL
(as shown in Table~\ref{tb:workloads}). CI represents the
computation-intensive workloads including kernels with high PUR,
whereas MI represents workloads with intensive memory accesses. MIX and ALL include a mix of
CI and MI kernels. ALL has more
kernels than MIX. In each workload, we assume the application
arrival conforms to Poisson distribution. The parameter $\lambda$
in the Poisson distribution affects the weight of the application in the
workload. For simplicity, we assume that all application has the
same $\lambda$. We also assume $\lambda$ is sufficiently large so
that at least two kernels are pending for execution at any time for
a high utilization of the GPU.

\begin{table}\scriptsize
\centering \caption{Workload configurations.}\label{tb:workloads}
\begin{tabular}
{|l|c|}
  \hline
  Workload & Applications            \\\hline
  CI       & BS, MM, TEA, MRIQ       \\\hline
  MI       & PC, SPMV, ST, SAD       \\\hline
  MIX      & PC, BS, TEA, SAD      \\\hline
  ALL      & PC, SPMV, ST, BS, MM, TE, MRIQ, SAD  \\\hline
\end{tabular}
\end{table}

\begin{figure*}
        \centering
        \begin{subfigure}{0.4\textwidth}
                \centering
                \includegraphics[width=\textwidth]{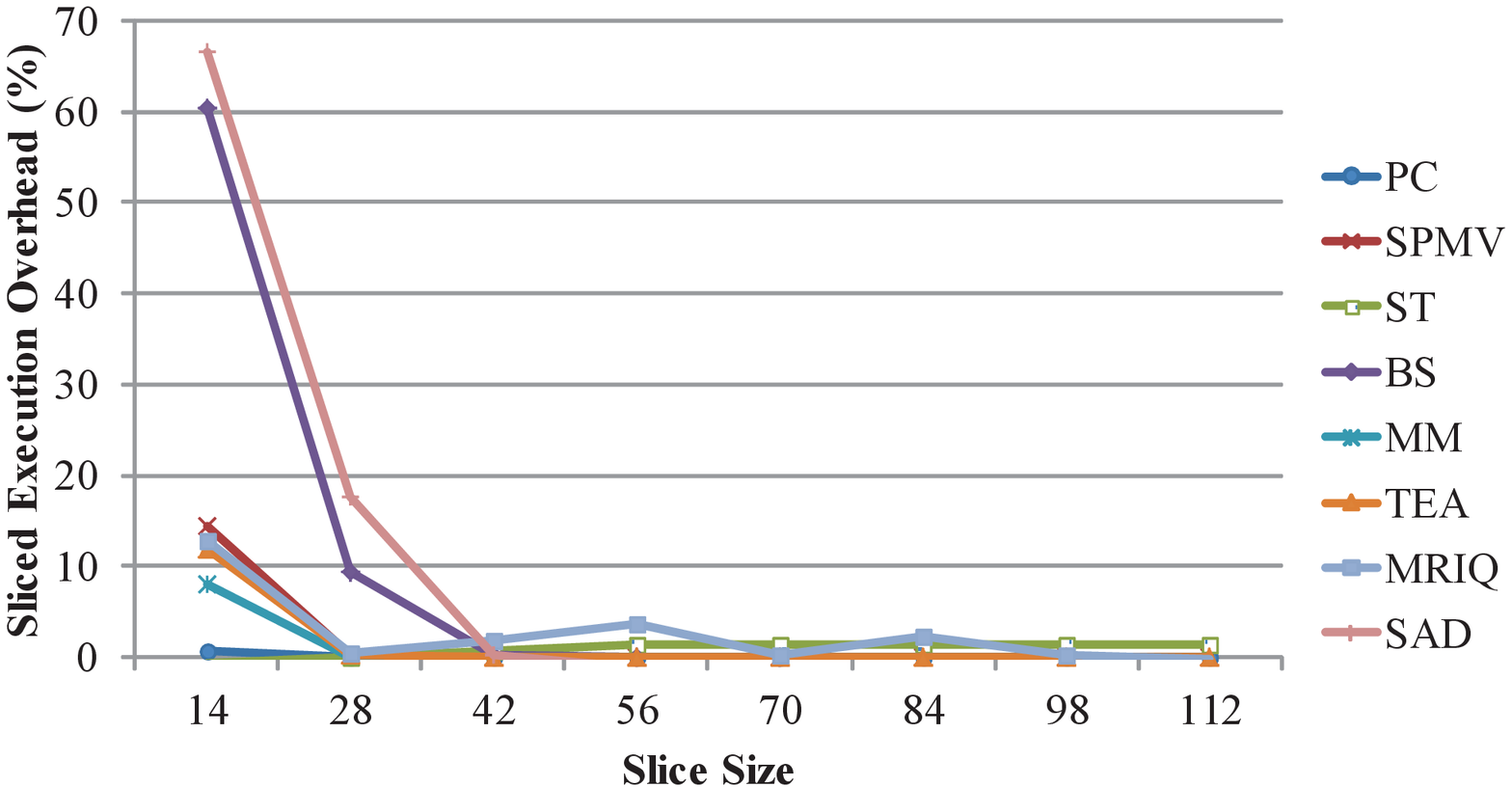}
                \vspace{-2ex}
                \caption{C2050}
                \label{fig:concurrent_estimation_c2050}
        \end{subfigure}
        \begin{subfigure}{0.4\textwidth}
                \centering
                \includegraphics[width=\textwidth]{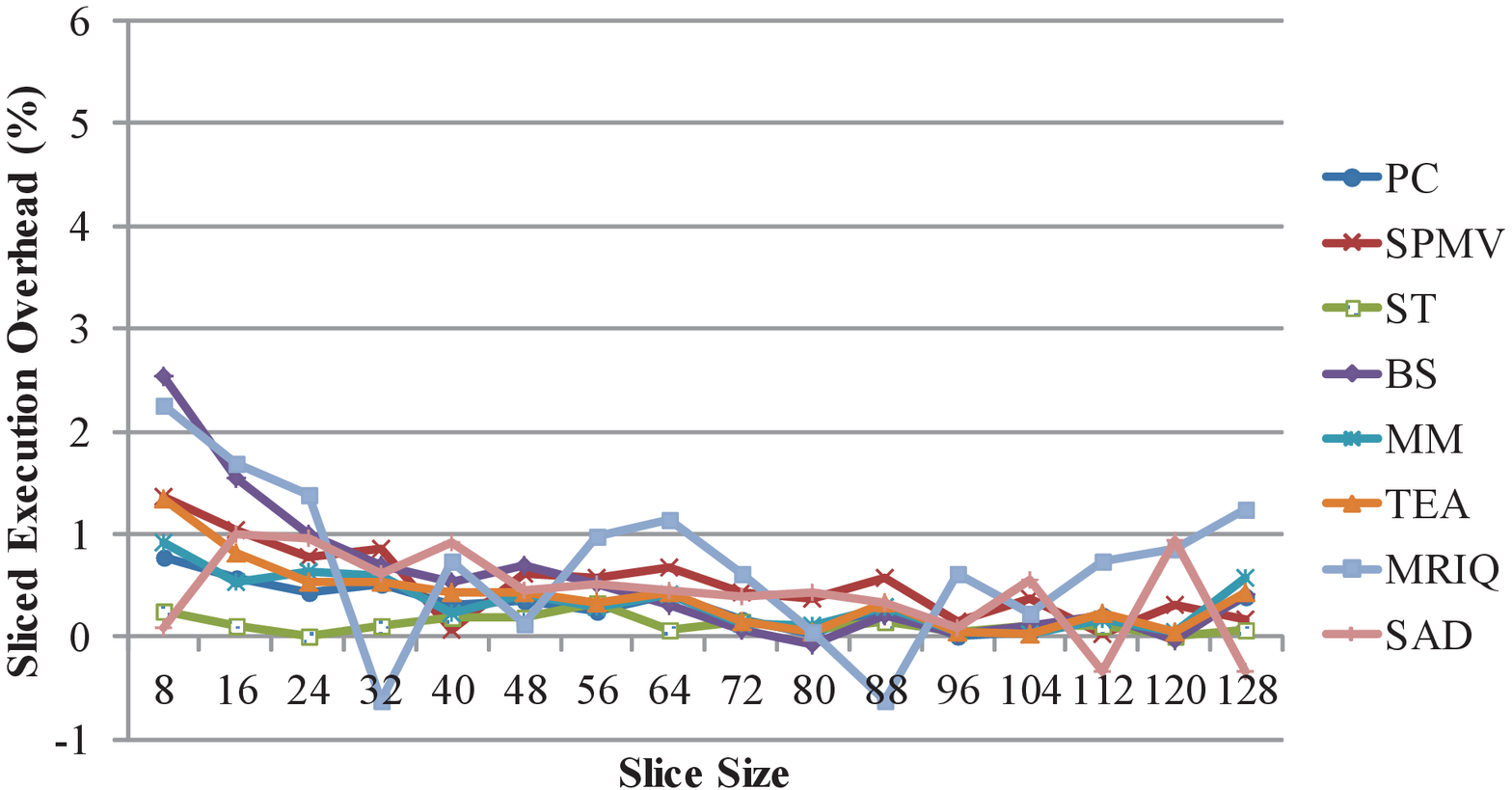}
                \vspace{-2ex}
                \caption{GTX680}
                \label{fig:concurrent_estimation_gtx680}
        \end{subfigure}
        \caption{Sliced execution overhead with varying slice size on both C2050 and GTX680.}
        \label{fig:slicingOverhead}
\end{figure*}

{\bf Comparisons.} To evaluate the effectiveness of kernel
scheduling in Kernelet, we have implemented the following scheduling
techniques:
\begin{itemize}
 \item
{\bf Kernel Consolidation (BASE)}: the kernel consolidation approach
of concurrent kernel execution~\cite{gpu_consolidation_hpdc11}.

\item
{\bf Oracle (OPT)}: OPT uses the same scheduling algorithm as Kernelet, except that it
pre-executes all possible slice ratios for all combinations to obtain the $\mathit{CP}$ and then determines the best slice ratio and kernel combination. In another word, OPT is an offline algorithm and provides the optimal throughput for the greedy scheduling algorithm.

\item
{\bf Monte Carlo-based co-schedule (MC)}: we develop a Monte Carlo
approach to generate the distribution of performance of different
co-schedule methods in the solution space. In each Monte Carlo
simulation, we randomly pick the kernel pairs and slice ratios for
co-scheduling. Through many Monte Carlo simulations, we can
quantitatively understand how different co-schedules affect the
performance. We denote the result of MC to be MC($s$), where $s$ is
the number of Monte Carlo simulations.
\end{itemize}

\subsection{Results on Kernel Slicing}
We first evaluate the overhead of sliced execution, which is defined
as $\frac{T_s}{T_{\mathit{ns}}}-1$, where $T_s$ and
$T_{\mathit{ns}}$ are the sliced and unsliced execution time,
respectively.

Figure~\ref{fig:slicingOverhead} shows the overhead for executing
individual kernels with varying slice sizes on C2050 and GTX680.
Slice sizes are set to multiples of the number of SMs on the GPU and
ranges from $|\mathit{SM}|$ to the maximum number \jianlong{under} the
occupancy limit. Overall, as the slice size increases, the slicing
overhead decreases. However, we observe quite different performance
behaviors on C2050 and GTX680, due to their architectural
differences. On C2050, when the size is small, the slicing overhead
is very high (up to 66.7\% for SAD). When the slice is larger than
or equal to 42 (three thread blocks per SM), the overhead is
ignorable for most kernels. Sliced execution overhead on GTX680 is
much smaller than on C2050. Almost all slice sizes lead to overhead
less than 2\% on GTX680. \jianlong{Regardless of the architectural
differences, the ignorable overhead of kernel slicing allows us to
exploit kernel slicing for co-scheduling slices from different
kernels with little additional cost.}

\subsection{Results on Model Prediction}
We evaluate the accuracy of our performance model in different
aspects, including the estimation of $\mathit{IPC}$s for single
kernels and concurrent kernel executions, and $\mathit{CP}$
prediction for concurrent kernel executions.

\begin{figure}
        \centering
        \begin{subfigure}{0.24\textwidth}
                \centering
                \includegraphics[width=\textwidth]{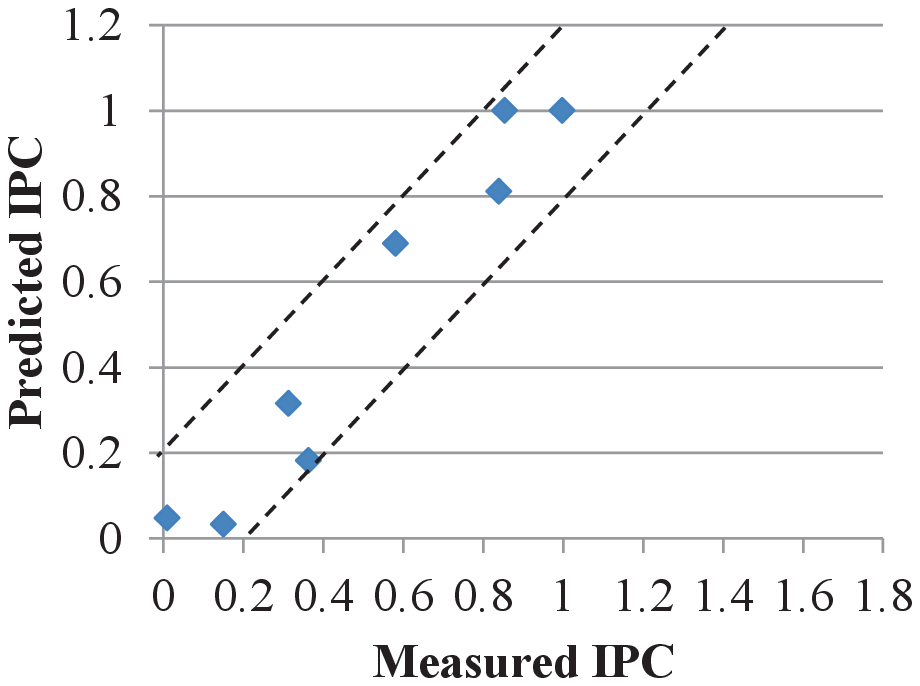}
                \vspace{-2ex}
                \caption{C2050}
                \label{fig:single_estimation_c2050}
        \end{subfigure}
        \begin{subfigure}{0.24\textwidth}
                \centering
                \includegraphics[width=\textwidth]{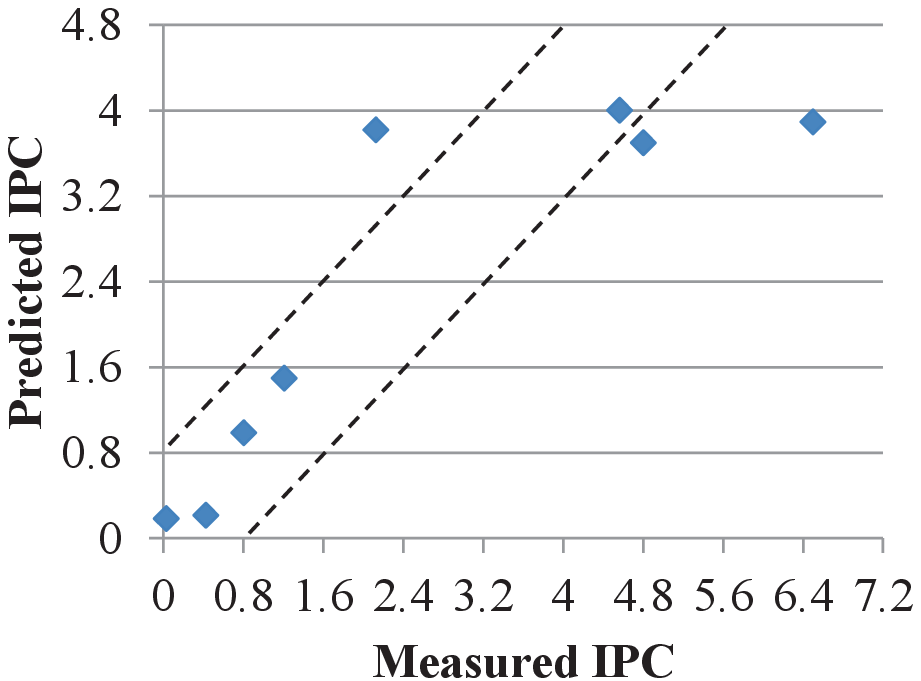}
                \vspace{-2ex}
                \caption{GTX680}
                \label{fig:single_estimation_gtx680}
        \end{subfigure}
        \caption{Comparison between predicted and measured single kernel execution IPCs on two GPUs.}
        \label{fig:singleKernelPrediction}
\end{figure}

{\bf Single Kernel Performance Prediction.}
Figure~\ref{fig:singleKernelPrediction} compares the measured and
estimated $\mathit{IPC}$ values for the eight benchmark applications
on C2050 and GTX680. We also show the two lines ( $y=x\pm 0.2$ for
C2050 and $y=x\pm 1.6$ for GTX680) to highlight the \jianlong{scope
where difference between measurement and estimation is within $\pm$
20\% of the peak IPC}. Note, the theoretical IPCs for C2050 and
GTX680 are one and eight respectively. If the result falls in this
scope, we consider the estimation well captures the trend of the
measurement. We can see that, most results are within the scope. We
further define the absolute error to be $\left\vert e-e'
\right\vert$, where $e$ and $e'$ are the measured and estimated
$\mathit{IPC}$ values, respectively. The average absolute error for
the eight benchmark applications is 0.08 and 0.21 on C2050 and
GTX680, respectively. Our probabilistic model has achieved a
reasonable accuracy in estimating the performance of single-kernel
executions on the GPU.

\begin{figure}
        \centering
        \begin{subfigure}{0.24\textwidth}
                \centering
                \includegraphics[width=\textwidth]{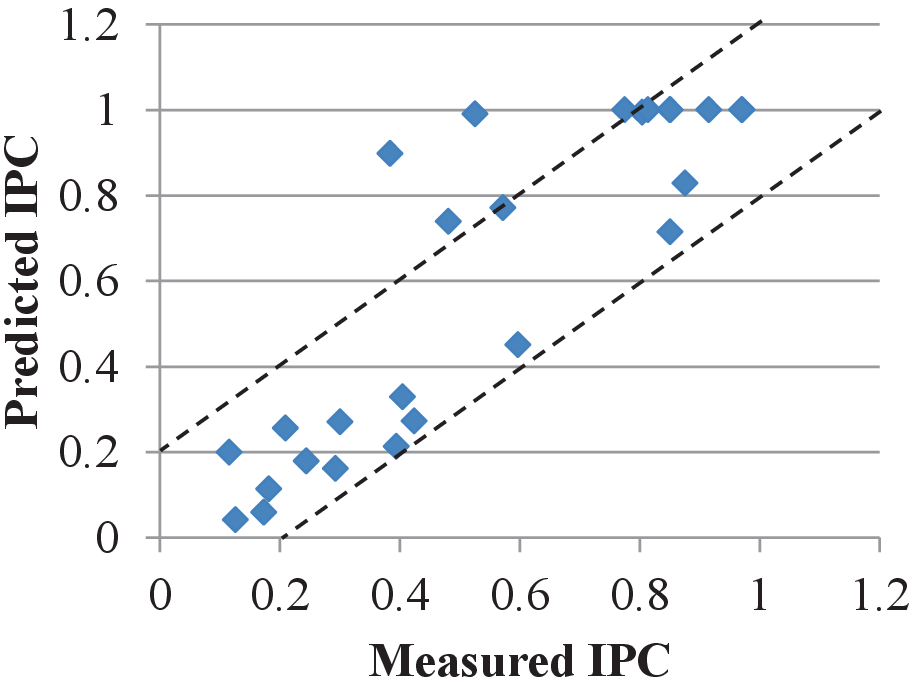}
                \vspace{-2ex}
                \caption{C2050}
                \label{fig:concurrent_estimation_c2050}
        \end{subfigure}%
        ~ 
        \begin{subfigure}{0.24\textwidth}
                \centering
                \includegraphics[width=\textwidth]{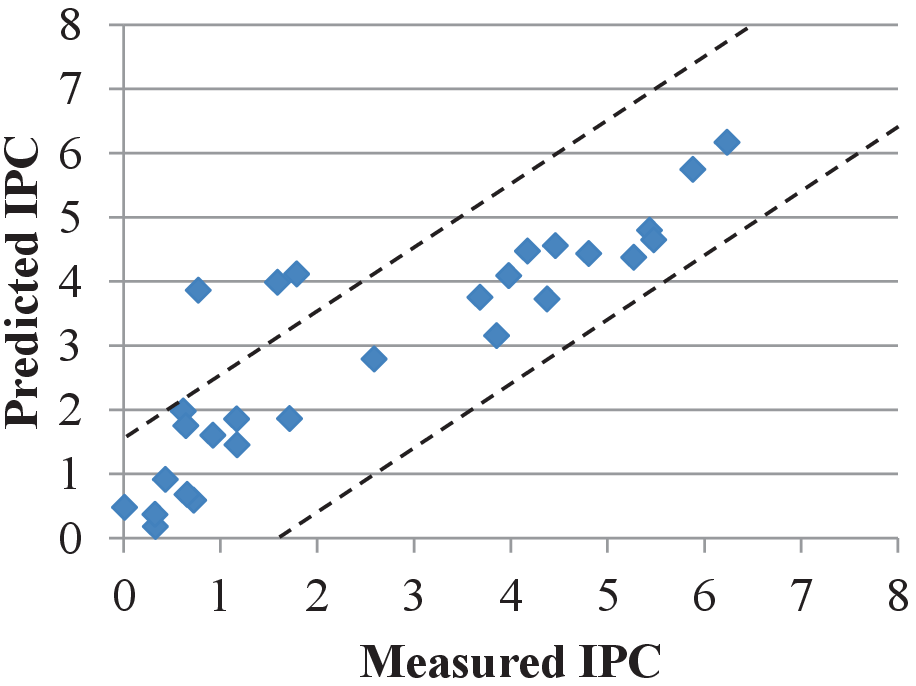}
                \vspace{-2ex}
                \caption{GTX680}
                \label{fig:concurrent_estimation_gtx680}
        \end{subfigure}

        \caption{Comparison between predicted and measured concurrent kernel execution IPCs on two GPUs with optimal slice ratio.}
        \label{fig:concurrentKernelPredictionOptimal}
\end{figure}

\begin{figure}
        \centering
        \begin{subfigure}{0.24\textwidth}
                \centering
                \includegraphics[width=\textwidth]{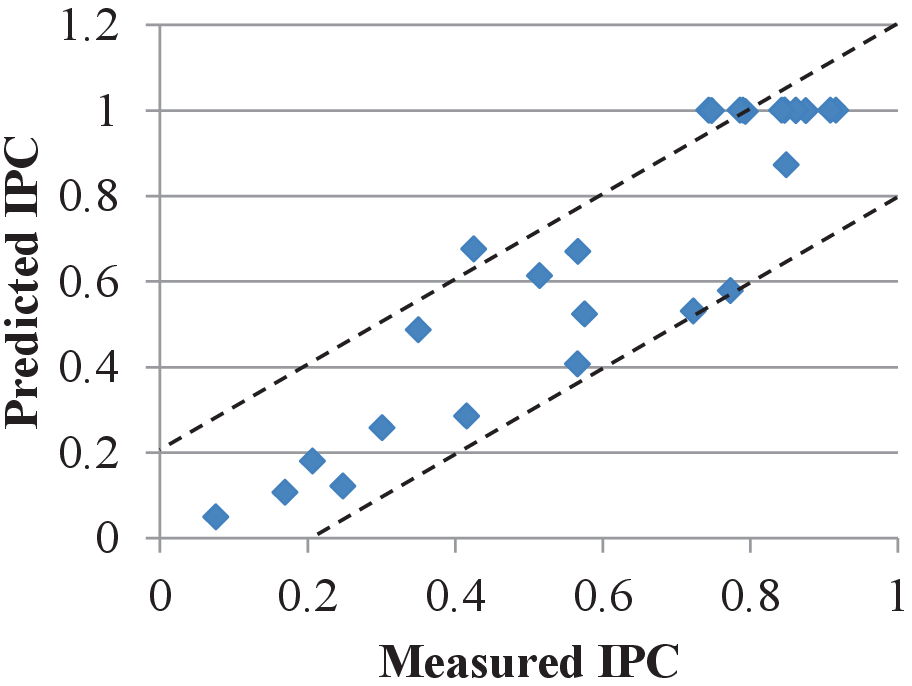}
                \vspace{-2ex}
                \caption{C2050}
                \label{fig:concurrent_estimation_onetoone_c2050}
        \end{subfigure}
        \begin{subfigure}{0.24\textwidth}
                \centering
                \includegraphics[width=\textwidth]{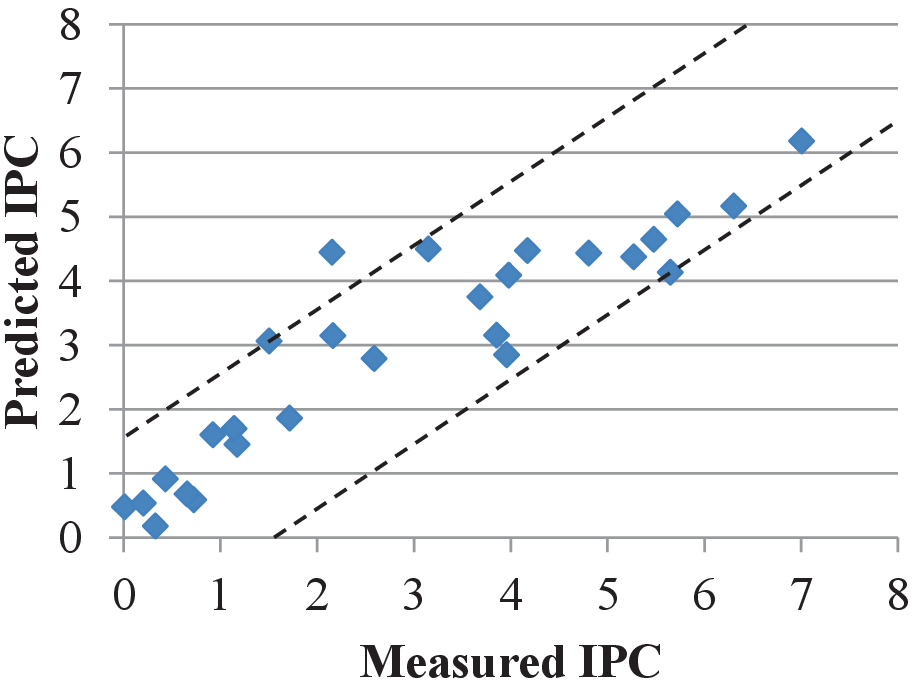}
                \vspace{-2ex}
                \caption{GTX680}
                \label{fig:concurrent_estimation_onetoone_gtx680}
        \end{subfigure}

        \caption{Comparison between predicted and measured concurrent kernel execution IPCs on two GPUs with fixed one-to-one slice ratio.}
        \label{fig:concurrentKernelPrediction121}
\end{figure}

{\bf Concurrent Kernel Performance Prediction.} For the eight
benchmark applications, we run every possible combination of kernel
pairs and measure the $\mathit{IPC}$ for each combination.
Figure~\ref{fig:concurrentKernelPredictionOptimal} compares the
measured and predicted $\mathit{IPCs}$ with the suitable slice ratio
given by our model. We have also studied other slicing ratios.
Figure~\ref{fig:concurrentKernelPrediction121} compares the measured
and predicted $\mathit{IPCs}$ with fixed $\mathit{ratio}=1:1$. We
observed similar results on other fixed ratios. Regardless of
different kernel combinations and slicing ratios, our model is able
to well capture the trend of concurrent executions for both dynamic
and static slice ratios.

\begin{figure}
  \centering
  \includegraphics[width=0.42\textwidth]{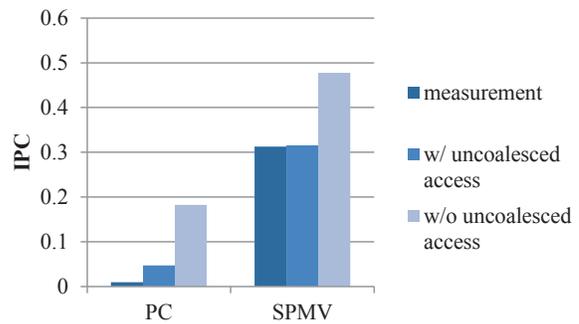}
  \caption{Comparison between predicted and measured concurrent kernel execution IPCs with/without \jianlong{considering uncoalesced access on C2050}.}
  \label{fig:concurrentKernelPredictionC2050NoUncoalesced}
\end{figure}

{\bf Model Optimizations.} We further evaluate the impact of
incorporating coalesced/uncoalesced memory accesses and the number
of warp schedulers on the GPU. Only two applications (PC and SPMV)
in our benchmark have uncoalesced memory accesses. We conduct the
single kernel execution prediction experiments by (wrongly) assuming
those two kernels with coalesced memory accesses only. The results
are shown in
Figure~\ref{fig:concurrentKernelPredictionC2050NoUncoalesced}.
Without considering uncoalesced access, the predicted IPC values are
much larger than measurements since the assumption of coalesced
access only underestimates the memory contention effects.

\begin{figure}
  \centering
  \includegraphics[width=0.30\textwidth]{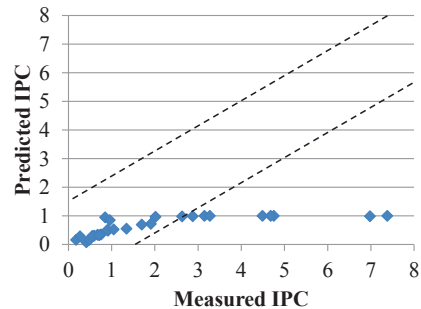}
  \caption{Comparison between predicted and measured concurrent kernel execution IPCs without considering multiple warp schedulers on GTX680.}
  \label{fig:concurrentKernelPredictionGTX680NoMutipipeline}
\end{figure}

Figure~\ref{fig:concurrentKernelPredictionGTX680NoMutipipeline}
shows the results of concurrent execution IPC prediction on GTX680 without considering the multiple warp schedulers. The estimation without considering the number of warp schedulers severely underestimates the IPC on GTX680, in comparison with the results in Figure~\ref{fig:concurrentKernelPredictionOptimal}.

\begin{figure}
  \centering
  \includegraphics[width=0.30\textwidth]{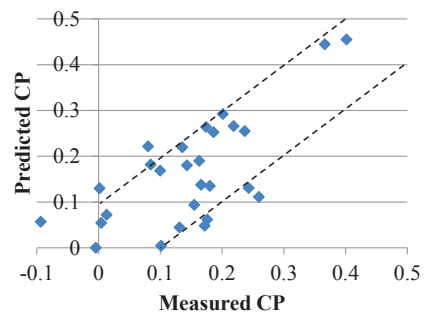}
  \caption{Comparison between predicted and measured ${CP}$ on C2050.}
  \label{fig:CPPrediction}
\end{figure}

{\bf $\mathit{CP}$ Prediction.} We further evaluate the accuracy of
$\mathit{CP}$ prediction. Figure~\ref{fig:CPPrediction} shows the
comparison between measured and predicted $\mathit{CP}$ on C2050. We
observe similar results on GTX680. The prediction is close to the
measurement. \jianlong{With accurate prediction on $\mathit{IPC}$,
the $\mathit{CP}$ difference between prediction and measurement is
small. The results are sufficiently good to guide the scheduling
decision as shown in the next section.}

\begin{figure*}
        \centering
        \begin{subfigure}{0.4\textwidth}
                \centering
                \includegraphics[width=\textwidth]{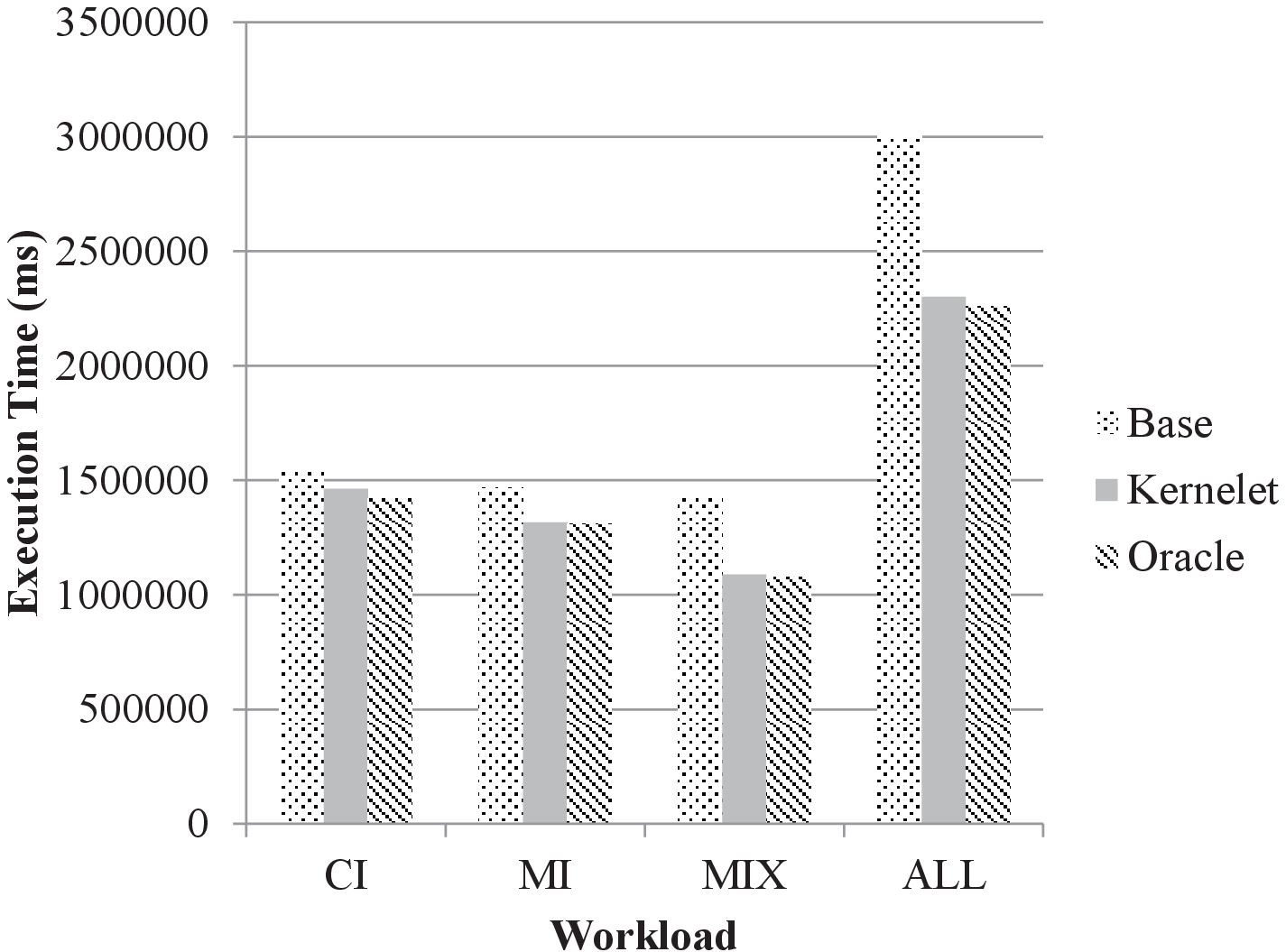}
                \vspace{-2ex}
                \caption{C2050}
                \label{fig:schedulingComparison_c2050}
        \end{subfigure}%
        ~ 
        \begin{subfigure}{0.4\textwidth}
                \centering
                \includegraphics[width=\textwidth]{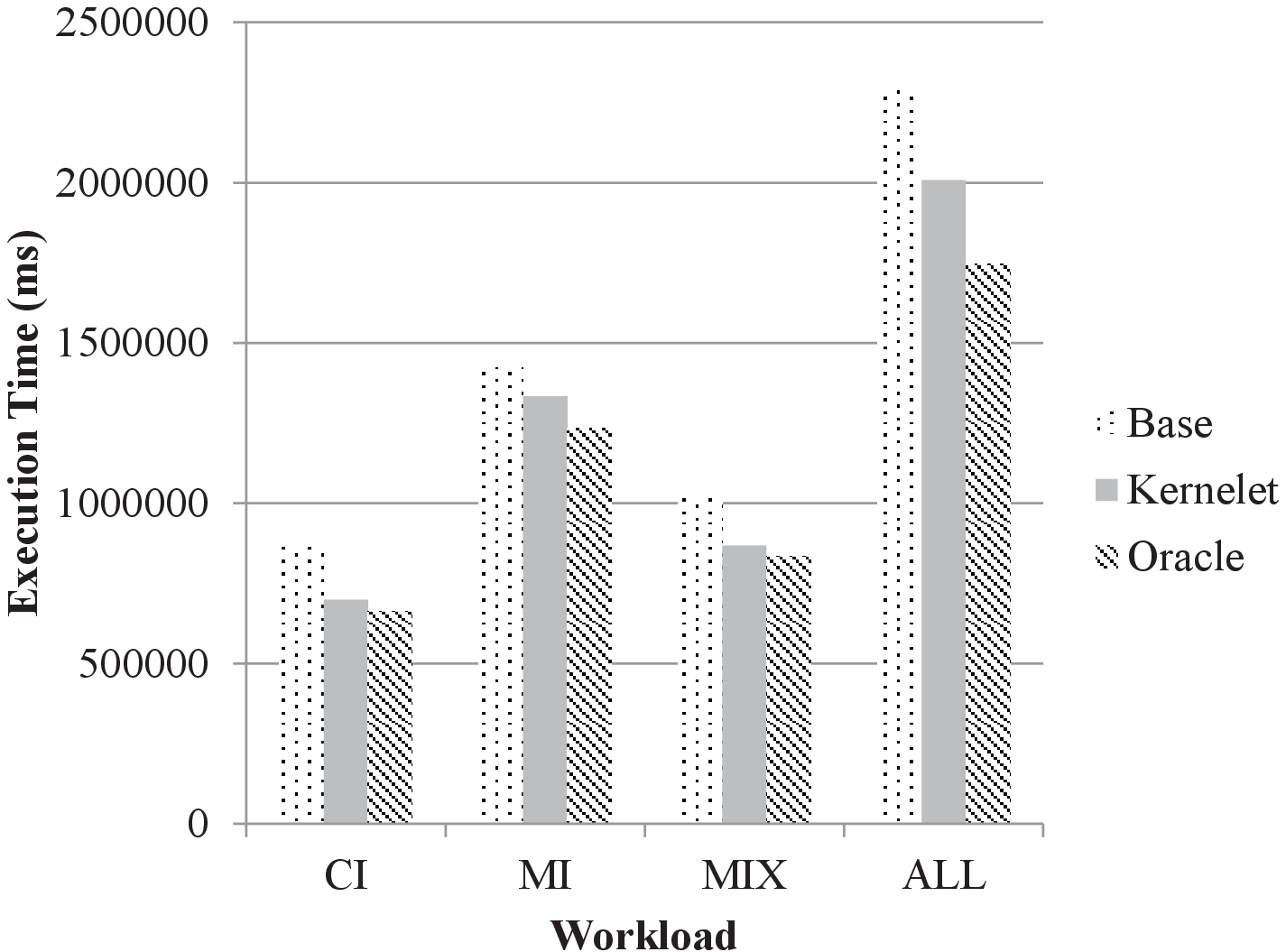}
                \vspace{-2ex}
                \caption{GTX680}
                \label{fig:schedulingComparison_gtx680}
        \end{subfigure}
        \caption{Comparison between different scheduling methods on both C2050 and GTX680.}
        \label{fig:schedulingComparison}
\end{figure*}

\subsection{Results on Kernel Scheduling}
In this section, we evaluate the effectiveness of our kernel
scheduling algorithm by comparing with BASE and OPT. To simulate the
continuous kernel submission process, we initiate 1000 instances for
each of each kernel mix and submit them for execution according to
Poisson distributions. Different scheduling algorithms are applied
and the total kernel execution time is reported.
Figure~\ref{fig:schedulingComparison} shows the total execution time
of those kernels on C2050 and GTX680. On all the four workloads with
different memory and computation characteristics, Kernelet
outperforms Base (with the improvement 5.0--31.1\% for C2050 and
6.7--23.4\% for GTX680). Kernelet achieves similar performance to
OPT (with the difference 0.7-3.1\% for C2050 and 4.0--15.0\% for
GTX680). The performance improvement of Kernelet over Base is more
significant on MIX and ALL, because \jianlong{Kernelet have more
chances to select kernel pairs with complementary resource usage}.
Still, Kernelet outperforms Base in CI and MI, because slicing
exposes the scheduling opportunities (even though they are small on
CI and MI).

Table~\ref{tb:pruningTableC2050} shows the number of kernels pruned
with different pruning parameters $\alpha_p$ and $\alpha_m$ on
C2050. Increasing $\alpha_p$ and $\alpha_m$ leads to more kernel
combinations being pruned. Similar pruning result is oberserved for
GTX680. Varying those two parameters can affect the pruning power
and also the optimization opportunities. Thus, we choose the default
values for $\alpha_p$ and $\alpha_m$ as 0.4 and 0.1 on C2050, and
0.4 and 0.105 on GTX680, respectively, as a tradeoff between pruning
power and optimization opportunities.

\begin{table}\scriptsize
\centering \caption{Number of kernels pruned with varying $\alpha_p$
and $\alpha_m$ on C2050.}\label{tb:pruningTableC2050}
\begin{tabular}
{|l|p{0.25cm}|p{0.25cm}|p{0.25cm}|p{0.25cm}|p{0.25cm}|p{0.25cm}|p{0.25cm}|p{0.25cm}|p{0.25cm}|p{0.25cm}|}

\hline

\backslashbox{$\alpha_m$}{$\alpha_p$} &0.1&0.2&0.3&0.4 & 0.5& 0.6& 0.7& 0.8& 0.9& 1.0\\\hline
0.015  & 0  & 0  & 2  & 2  & 3  & 4  & 4  & 4  & 4  & 4  \\\hline
0.03  & 0  & 2  & 5  & 6  & 7  & 8  & 9  & 9  & 9  & 9  \\\hline
0.045  & 0  & 3  & 7  & 8  & 9  & 10  & 11  & 11  & 11  & 11  \\\hline
0.06  & 1  & 4  & 8  & 9  & 12  & 13  & 15  & 15  & 15  & 15  \\\hline
0.075  & 1  & 4  & 8  & 9  & 12  & 13  & 15  & 15  & 15  & 15  \\\hline
0.09  & 1  & 4  & 8  & 9  & 12  & 13  & 15  & 15  & 16  & 16  \\\hline
0.105  & 1  & 4  & 9  & 10  & 14  & 15  & 18  & 18  & 20  & 20  \\\hline
0.12  & 2  & 6  & 11  & 12  & 17  & 18  & 21  & 22  & 24  & 24  \\\hline
0.135  & 2  & 6  & 11  & 12  & 17  & 19  & 22  & 23  & 25  & 26  \\\hline
0.15  & 2  & 6  & 11  & 13  & 18  & 20  & 23  & 24  & 27  & 28  \\\hline

\end{tabular}
\end{table}

\begin{figure}
  \centering
  \includegraphics[width=0.36\textwidth]{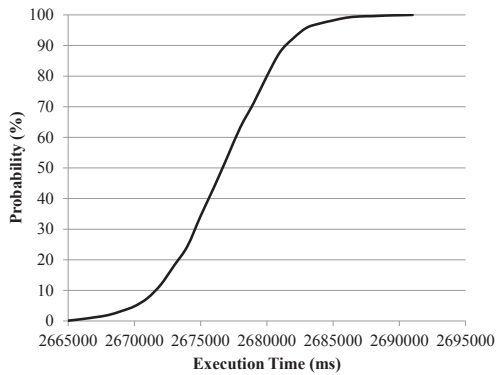}
  \caption{CDF (cumulative distribution function) of execution time of MC($1000$).}
\label{fig:randomSchedulingDistribution}
\end{figure}

We finally study the execution time distribution of the scheduling
candidate space. Figure~\ref{fig:randomSchedulingDistribution} shows
the CDF (cumulative distribution function) of the execution time of
the MC($1000$). As we can see from the figure, none of the random
schedules is better than Kernelet. It demonstrates that random
co-schedules hurt the performance in a high probability due to the
huge space of schedule plans.

\section{Related Work}\label{sec:related}
In this section, we review the related work in two categories: 1)
scheduling algorithms on CPUs, especially for CPUs with SMT
(Simultaneous Multi-Threading), and 2) multi-kernel executions on
GPUs.

\subsection{CPU Scheduling and Performance Modeling}
SMT has been an effective hardware technology to better utilize CPU
resources. SMT allows instructions from multiple threads to be
executed on the instruction pipeline at the same time.
\jianlong{Various SMT aware scheduling techniques have been proposed
to increase the CPU
utilization\cite{SMTThreadSensitive,SMTSymbiotic,SMTSymbioticPriority,SMTModelContentionICCD05,SMTClustering,SMTSymbiosis}}.
The core idea of SMT aware scheduling is to co-schedule threads with
complementary resource requirements. Several mechanisms have been
proposed to select the threads to be executed on the same SMT core,
such as hardware counters feedback~\cite{SMTThreadSensitive,
SMTModelContentionICCD05}, pre-execution of all
combinations~\cite{SMTSymbiotic} and probabilistic job symbiosis
model~\cite{SMTSymbiosis}. Performance models including the Markov
chain based ones have also been adopted for concurrent tasks
modeling on the CPUs. Serrano et
al.~\cite{model_multistreamed_superscalar,
performance_estimation_smt} developed a model to estimate the
instruction throughput of super-scalar processors executing multiple
instruction streams. Chen et al.~\cite{first_order_model} proposed
an analytical model for estimating throughput of multi-threaded
CPUs.

Despite the fruitful research results on SMT scheduling, they are
not applicable to GPUs due to the architecture differences. First,
main performance affecting issues are different for CPU and GPU
applications. L2 data cache is a key consideration issue for
SMT-aware thread scheduling, whereas thread parallelism is usually
more important for the performance of GPGPU
programs~\cite{saramodel_ppopp10, kimmodel_isca09,
kimmodel_ppopp12}. Second, scheduling on GPUs is not as flexible as
that on CPUs. Current GPUs do not support task preemption. Third,
unlike CPUs supporting the concurrent execution of a relatively
small number of threads, each GPGPU kernel launches thousands of
threads. Additionally, the maximum number of co-scheduling threads
equals to the number of hardware context on the CPU, while the
number of active warps on GPUs is dynamic, depending on the resource
usage of the thread blocks. The slicing, scheduling and
performance models in Kernelet are specifically designed for GPUs, taking those
issues into consideration.

\subsection{GPU Multiple Kernel Execution and Sharing}
In the past few years, GPU architectures have undergone significant
and rapid improvements for GPGPU support. Due to lack of concurrent
kernel support in early GPU architectures, researchers initially proposed
to merge two kernels at the source code
level~\cite{cmerge_cuda_scheduler,cmerge_hotpar12}. In those
methods, two kernels are combined into a single kernel with if-else
branches on different granularities (e.g., thread blocks). They have
three major disadvantages compared with our approach. First, combining
the code of two kernels will increase the resource usage of each
thread block, leading to lower SM occupancy and
performance degradation~\cite{cudaguide}. Second, those approaches
require source code, which may not be always available in the shared
environments. Third, it requires
two kernels with different block sizes avoiding using barriers
within the thread block, otherwise deadlock may occur.

Recently, new-generation GPUs like NVIDIA Fermi GPUs support
concurrent kernel executions. Taking advantage of this new
capability, a number of multi-kernel optimization
techniques~\cite{ptask_sosp11, GVim, gpusharing_icpp11} have been
developed to improve the utilization of GPUs. Ravi et
al.~\cite{gpu_consolidation_hpdc11} proposed kernel consolidation to
enable space sharing \jianlong{(different kernels run on different
SMs) and time sharing (multiple kernels reside on the same SM)} on
GPUs. Space sharing happens when the total number of thread blocks
of all kernels does not exceed the number of SMs and each block can
be executed on a dedicated SM. If the total number of thread blocks
is larger than the number of SMs, while SMs have sufficient
resources to accommodate more thread blocks from different kernels,
time sharing happens. That means, kernel consolidation does not have
space sharing and have little time sharing when the launched kernels
have sufficient thread blocks to occupy the GPU. Furthermore, they
determined the kernel to be consolidated with heuristics based on
the number of thread blocks. In contrast, Kernelet utilizes slicing
to create more opportunities for time sharing, and develops a
performance model to guide the scheduling decision. Peters et
al.~\cite{persistent_kernel_cit_10} used a persistently running
kernel to handle requests from multiple applications. \jianlong{GPU
virtualization has also been investigated~\cite{GVim,
gpusharing_icpp11}}.

Recent studies also address the problem of GPU scheduling when
multiple users co-reside in one machine.
Pegasus~\cite{pegasus_atc11} coordinates computing resources like
accelerators and CPU and provides a uniform resource usage model.
Timegraph~\cite{timegraph_atc11} and PTask~\cite{ptask_sosp11}
manage the GPU at the operating system level. Kato et
al.~\cite{rgem_rtss11} introduced the responsive GPGPU execution
model (RGEM). All those scheduling methods do not consider how to
schedule concurrent kernels in order to fully utilize the GPU
resources.

As for performance models on GPUs, Hong~\cite{kimmodel_isca09} and
Kim~\cite{kimmodel_ppopp12} proposed analytical models based on the
round-robin warp scheduling assumption. Baghsorkhi et
al.~\cite{saramodel_ppopp10} introduced the work flow graph
interpretation of GPU kernels to estimate their execution time. All
those models are designed for a single kernel. Moreover, they
usually require extensive hardware profiling and/or simulation
processes. In contrast, our performance model is designed for
concurrent kernel executions on the GPU, and relies on a small set
of key performance factors of individual kernel to predict the
performance of concurrent kernel executions.

\section{Conclusion}\label{sec:conclusion}

Recently, GPUs have been more and more widely used in clusters and
cloud environments, where many kernels are submitted and executed on
the shared GPUs. This paper proposes Kernelet to improve the
throughput of concurrent kernel executions for such shared
environments. Kernelet creates more sharing opportunities with
kernel slicing, and uses a probabilistic performance model to
capture the non-deterministic performance features of
multiple-kernel executions. We evaluate Kernelet on two NVIDIA GPUs,
Tesla C2050 and GTX680, with Fermi and Kepler architectures
respectively. Our experiments demonstrate the accuracy of our
performance model, and the effectiveness of Kernelet by improving
the concurrent kernel executions by 5.0--31.1\% and 6.7--23.4\% on
C2050 and GTX680 on our workloads, respectively.

\bibliographystyle{abbrv}
\begin{small}

\end{small}

\end{document}